\documentclass[final,journal,letterpaper,twoside,twocolumn]{IEEEtran}

\usepackage[utf8]{inputenc} 
\usepackage[T1]{fontenc} 
\usepackage[cmex10]{amsmath} 

\usepackage{amssymb} 
\usepackage{mathtools} 
\usepackage{amsfonts} 
\usepackage{accents} 

\usepackage{mleftright}\mleftright 

\usepackage{multirow}

\usepackage[dvips]{graphicx} 
\graphicspath{{./}}
\DeclareGraphicsExtensions{.eps}

\usepackage[dvipsnames,table]{xcolor} 

\usepackage{bbm} 
\usepackage{xfrac} 
\usepackage{cancel} 
\usepackage{wrapfig} 
\usepackage{tensor} 

\usepackage{theorem}
\usepackage{cite}

\usepackage{url} 

\usepackage{psfrag} 

\usepackage[inline]{enumitem} 

\usepackage{array} 

\usepackage{aliascnt} 

\usepackage[hidelinks,linktocpage]{hyperref}

\usepackage{cleveref} 

\usepackage{orcidlink} 

\usepackage[ruled,algoruled,vlined]{algorithm2e}

\SetCommentSty{mycommfont}

\SetKw{Breakwhile}{break while}
\SetKw{Breakfor}{break for}
\SetKw{Break}{break}

\usepackage{tikz}
\usetikzlibrary{positioning,arrows,decorations.pathmorphing}
\usetikzlibrary{decorations.pathreplacing}
\usetikzlibrary{shapes,calc}
\usetikzlibrary{quotes,bayesnet}

\usepackage{xparse} 

\usepackage{ifthen} 

\usepackage{comment}

\crefname{section}{Section}{Sections}
\crefname{subsection}{Section}{Sections}
\crefname{equation}{Eq.}{Equations}
\crefname{enumi}{part}{parts}
\crefname{table}{Table}{Tables}
\crefname{figure}{Figure}{Figures}
\crefname{algocf}{Algorithm}{Algorithms}

\newtheorem{theorem}{Theorem}
\crefname{theorem}{Theorem}{Theorems}

\newaliascnt{lemma}{theorem}
\newtheorem{lemma}[lemma]{Lemma}
\aliascntresetthe{lemma}
\crefname{lemma}{Lemma}{Lemmas}

\newaliascnt{definition}{theorem}
\newtheorem{definition}[definition]{Definition}
\aliascntresetthe{definition}
\crefname{definition}{Definition}{Definitions}

\newaliascnt{corollary}{theorem}
\newtheorem{corollary}[corollary]{Corollary}
\aliascntresetthe{corollary}
\crefname{corollary}{Corollary}{Corollarys}

\newaliascnt{claim}{theorem}

\aliascntresetthe{claim}
\crefname{claim}{Claim}{Claims}

\newaliascnt{conjecture}{theorem}

\aliascntresetthe{conjecture}
\crefname{conjecture}{Conjecture}{Conjectures}

\newaliascnt{question}{theorem}

\aliascntresetthe{question}
\crefname{question}{Question}{Questions}

\newaliascnt{example}{theorem}
\newtheorem{example}[example]{Example}
\aliascntresetthe{example}
\crefname{example}{Example}{Examples}

\newaliascnt{oquestion}{theorem}

\aliascntresetthe{oquestion}
\crefname{oquestion}{Open Question}{Open Questions}

\theoremstyle{plain}
\theorembodyfont{\normalfont}

\newaliascnt{remark}{theorem}

\aliascntresetthe{remark}
\crefname{remark}{Remark}{Remark}

\newtheorem{cnstr}{Construction}

\crefname{cnstr}{Construction}{Constructions}

\crefname{step}{Step}{Steps}

\crefname{regime}{Regime}{Regimes}

\newtheorem{myalgo}{Algorithm}

\crefname{myalgo}{Algorithm}{Algorithms}

\newcommand\numberthis{\stepcounter{equation}\tag{\theequation}}

\newcounter{enumrom}
\renewcommand{\theenumrom}{(\roman{enumrom})}

\makeatletter
\renewcommand{\@endtheorem}{\endtrivlist}
\makeatother

\makeatletter
\renewcommand{\thefigure}{{\@arabic\c@figure}}
\renewcommand{\fnum@figure}{{\bf Figure\,\thefigure}}
\makeatother

\renewcommand{\leq}{\leqslant}
\renewcommand{\geq}{\geqslant}

\newcommand{\bfe}{{\boldsymbol e}}

\newcommand{\bfs}{{\boldsymbol s}}

\newcommand{\bfu}{{\boldsymbol u}}

\newcommand{\bfx}{{\boldsymbol x}}
\newcommand{\bfy}{{\boldsymbol y}}
\newcommand{\bfz}{{\boldsymbol z}}
\newcommand{\bfA}{{\mathbf A}}

\newcommand{\bfC}{{\mathbf C}}

\newcommand{\bfG}{{\mathbf G}}

\newcommand{\bfT}{{\mathbf T}}

\newcommand{\bfzero}{{\boldsymbol 0}}

\newcommand{\cA}{\mathcal{A}}

\newcommand{\cC}{\mathcal{C}}

\newcommand{\cF}{\mathcal{F}}

\newcommand{\cH}{\mathcal{H}}

\newcommand{\cR}{\mathcal{R}}

\renewcommand{\Bbb}{\mathbb}

\newcommand{\N}{{\Bbb N}}

\newcommand{\Z}{{\Bbb Z}}

\DeclarePairedDelimiter\abs{\lvert}{\rvert}
\DeclarePairedDelimiter\ceilenv{\lceil}{\rceil}
\DeclarePairedDelimiter\floorenv{\lfloor}{\rfloor}
\DeclarePairedDelimiter\parenv{\lparen}{\rparen}

\DeclarePairedDelimiter\bracenv{\lbrace}{\rbrace}
\DeclarePairedDelimiterX\mathset[2]{\lbrace}{\rbrace}{#1 : #2}
\DeclarePairedDelimiterX\inner[2]{\langle}{\rangle}{#1 \mathrel{},\mathrel{} #2}
\DeclarePairedDelimiterX\condparenv[2]{(}{)}{#1 \mathrel{}\delimsize\vert\mathrel{} #2}

\newcommand{\nhphantom}[2]{\sbox0{$\left#1\vphantom{#2}\right.$}\hspace{-0.58\wd0}}
\DeclarePairedDelimiterX{\multbrace}[1]{\lbrace}{\rbrace}{
    \nhphantom{\lbrace}{#1} \delimsize\lbrace \mathopen{} #1 \mathclose{} \delimsize\rbrace \nhphantom{\rbrace}{#1}
}

\DeclarePairedDelimiterX{\multset}[2]{\lbrace}{\rbrace}{
    \nhphantom{\lbrace}{#1 : #2} \delimsize\lbrace \mathopen{} #1 : #2 \mathclose{} \delimsize\rbrace \nhphantom{\rbrace}{#1 : #2}
}

\DeclareDocumentCommand\norm{ o m }{
    \IfNoValueTF{#1}
        {\left\Vert#2\right\Vert}
        {\left\Vert#2\right\Vert_{#1}}
}

\DeclareDocumentCommand\der{ o m o }{
    \IfNoValueTF{#1}
        {
            \IfNoValueTF{#3}
                {\frac{d}{d{#2}}}
                {\frac{d{#3}}{d{#2}}}
        }
        {\parenv*{\frac{d}{d{#2}}}^{#1}\IfNoValueTF{#3}{}{#3}}
}
\DeclareDocumentCommand\partder{ o m m }{
    \IfNoValueTF{#1}
        {\frac{\partial{#3}}{\partial{#2}}}
        {\frac{\partial^{#1}{#3}}{{\partial{#2}}^{#1}}}
}
\DeclareDocumentCommand\df{ o m o }{
    d\IfNoValueTF{#1}{}{^{#1}}{#2}\IfNoValueTF{#3}{}{_{#3}}
}

\newcommand{\deq}{\mathrel{\triangleq}}

\newcommand{\rv}{^{\mathrm{r}}}

\newcommand{\tildel}{\widetilde{\delta}_q}
\newcommand{\pid}{\pi_{\mathrm{d}}}
\newcommand{\Hq}{\bar{H}_q}

\DeclareMathOperator{\supp}{supp}

\DeclareMathOperator{\wt}{wt}

\DeclareMathOperator{\red}{red}

\DeclareMathOperator{\rf}{\cR\cF}
\DeclareMathOperator{\rrf}{\cR\cR\cF}
\DeclareMathOperator{\ha}{\cH\cA}
\DeclareMathOperator{\en}{en}

\DeclareDocumentCommand\enc{ o }{
    \IfNoValueTF{#1}
        {\operatorname{Enc}}
        {\operatorname{Enc}_{\ref*{#1}}}
}

\DeclareDocumentCommand\dec{ o }{
    \IfNoValueTF{#1}
        {\operatorname{Dec}}
        {\operatorname{Dec}_{\ref*{#1}}}
}

\makeatletter
\newcommand\code[1]{%
  \@ifundefined{r@#1}{%
    \cC_{\operatorname*{#1}}%
  }{%
    \cC_{\ref*{#1}}%
  }%
}
\makeatother

\DeclareDocumentCommand\ball{ o }{
    \IfNoValueTF{#1}
        {B}
        {B^{\mathrm{#1}}}
}
\DeclareDocumentCommand\sphere{ o }{
    \IfNoValueTF{#1}
        {S}
        {S^{\mathrm{#1}}}
}

\newcommand{\ds}{d_{\mathrm{H}}}

\newcommand{\rrfs}{\rrf^{\mathrm{s}}}

\newcommand{\rrfd}{\rrf^{\mathrm{d}}}

\begin{document}

\title{On Codes for the Noisy Substring Channel}

\author{%
  Yonatan~Yehezkeally\,\orcidlink{0000-0003-1652-9761}\,%
		,~\IEEEmembership{Member,~IEEE}
  and Nikita~Polyanskii\,\orcidlink{0000-0003-3735-5705}\,%
       ,~\IEEEmembership{Member,~IEEE}
  \thanks{%
  Manuscript received 25~September~2023; revised 6~February~2024; accepted 24~March~2024. 
  This work has received funding from the European Research Council (ERC) under the European Union’s Horizon 2020 research and innovation programme (Grant agreement No. 801434). 
  The work of Yonatan~Yehezkeally was supported by the Alexander von Humboldt Foundation under a Carl Friedrich von Siemens Post-Doctoral Research Fellowship. 
  The work of Nikita~Polyanskii was supported by the German Research Foundation (Deutsche Forschungsgemeinschaft, DFG) under Grant No. WA3907/1-1.
  An earlier version of this paper was presented in part at the 2021 {IEEE} International Symposium on Information Theory ({ISIT}) [\textsc{DOI:\; 10.1109/ISIT45174.2021.9517943}]. 
  \emph{(Corresponding author: Yonatan Yehezkeally.)}}
  \thanks{%
  Yonatan~Yehezkeally is with the Institute for Communications Engineering, School of Computation, Information and Technology, Technical University of Munich (TUM), 80333 Munich, Germany 
  (e-mail: \texttt{yonatan.yehezkeally@tum.de}).
  Nikita~Polyanskii is with IOTA Foundation, Berlin, Germany.}
  \thanks{Copyright (c) 2024 IEEE. Personal use of this material is permitted. Permission from IEEE must be obtained for all other uses, in any current or future media, including reprinting/republishing this material for advertising or promotional purposes, creating new collective works, for resale or redistribution to servers or lists, or reuse of any copyrighted component of this work in other works.}
}

\maketitle

\begin{abstract}
We consider the problem of coding for the substring channel, in which 
information strings are observed only through their (multisets of) 
substrings. Due to existing DNA sequencing techniques and applications 
in DNA-based storage systems, interest in this channel has renewed in 
recent years. 
In contrast to existing literature, we consider a noisy channel model 
where information is subject to noise \emph{before} its substrings are 
sampled, motivated by in-vivo storage.

We study two separate noise models, substitutions or deletions. In 
both cases, we examine families of codes which may be utilized for 
error-correction and present combinatorial bounds on their sizes. 
Through a generalization of the concept of repeat-free strings, we 
show that the added required redundancy due to this imperfect 
observation assumption is sublinear, either when the fraction of 
errors in the observed substring length is sufficiently small, or when 
that length is sufficiently long. This suggests that no asymptotic 
cost in rate is incurred by this channel model in these cases. 
Moreover, we develop an efficient encoder for such constrained strings 
in some cases.

Finally, we show how a similar encoder can be used to avoid formation 
of secondary-structures in coded DNA strands, even when accounting for 
\emph{imperfect} structures.
\end{abstract}


\section{Introduction}

\IEEEPARstart{D}{{NA}} 
as a medium for data storage offers high density and longevity, 
far greater than those of electronic media \cite{ChuGaoKos12}. Among 
its applications, data storage in DNA may offer a protected medium for 
long-period data storage \cite{Bal13,WonWonFoo03}. 
In particular, it has recently been demonstrated that storage in the 
DNA of living organisms (henceforth, \textit{in-vivo} DNA storage) is 
now feasible \cite{Shi17}; the envelope of a living cell affords some 
level of protection to the data, and even offers propagation, through 
cell replication. Among its varied usages, in-vivo DNA storage allows 
watermarking genetically modified organisms (GMOs) \cite{AriOha04,
HeiBar07,LisDauBruKliHamLeiWag12} to protect intellectual property, 
or labeling research material \cite{WonWonFoo03,JupFicSamQinFig10}. 
It may even conceal sensitive information, as it may appear 
indistinguishable from the organism's own genetic information 
\cite{CleRisBan99}.

Similarly to other media, information stored over this medium is 
subjected to noise due to mutations, creating errors in data, which 
accumulate over time and replication cycles. Examples of such noise 
include symbol insertions or deletion, in addition to substitutions 
(point-mutations) \cite{HecMikGra19,SabOrlShaAnaYaaYak21}; 
the latter is the focus of the vast majority of 
classical error-correction research, and the former have also been 
studied. Interestingly, however, the very methods we currently use to 
store and later retrieve data from DNA inherently introduce new 
constraints on information reconstruction. 
While desired sequences may be synthesized (albeit, while suffering 
from errors, e.g., substitution noise), the process of DNA sequencing, 
i.e., retrieving the DNA sequence of an organism, only observes that 
sequence as the (likely incomplete) multiset of its substrings 
(practically, up to a certain substring length) \cite{KiaPulMil16}. 
Thus, information contained in the order of these substrings might be 
irrevocably lost. As a result of these constraints, conventional and 
well-developed error-correction approaches cannot simply be applied.

To overcome these effects, one approach in existing literature is to 
add redundancy in the form of indexing, in order to recover the order 
of substrings (see, e.g., \cite{LenSieWacYaa20,SimRavBru20,
SimRavBru21}). A different approach, potentially more applicable to 
in-vivo DNA storage, is to add redundancy in the form of constraints 
on the long information string, such that it can be uniquely 
reconstructed by knowledge of its substrings of a given length (or 
range of lengths). 
The combinatorial problem of recovering a sequence from its substrings 
has attracted attention in recent years \cite{Ukk92,AchDasMilOrlPan15,
ShoCouTse16,ChaChrEzeKia17,GabMil19,EliGabMedYaa21,MarYaa21,
ChrKiaRaoVarYaaYao23}, and coding schemes involving only these substrings 
(including the incidence frequency of each substring) were studied 
\cite{KiaPulMil16,LenSieWacYaa19,RavSchYaa19,SimRavBru21,BeeSch22}. 

However, works dedicated to overcoming this obstacle, inherent to the 
technology we use, have predominantly focused on storage outside of 
living cells (i.e., \emph{in-vitro} DNA storage). Likewise, works 
focused on error-correction for in-vivo DNA data storage (e.g., 
\cite{JaiFarSchBru17b,AloBruFarJai17,FarSchBru19}) have disregarded 
the technical process by which data is to be read. 
However, in real applications varied distinct noise mechanisms act on 
stored data concurrently. Hence, in practice, both sets of challenges 
have to be collectively overcome in order to robustly store 
information using in-vivo DNA.

The aim of this work is to protect against errors in the information 
string (caused by mutations over the replication process of cells), 
when channel outputs are given by the multisets of their substrings, 
of a predetermined length, rather than entire strings. This models the 
process of DNA sequencing, once information needs to be read from the 
medium. We shall study the required redundancy of this model, and 
devise coding strategies, under the assumption of two different error 
types: substitution and deletion noise.

Another application for this line of research is secondary-structure 
avoidance. Secondary structures are complex spatial structures that 
can form in a chemically active single-stranded DNA, as a result of 
the strand folding upon itself to allow two sub-segments to bond via 
complementary-base-pair hybridization \cite{MilKas06}. Their formation 
renders the DNA strand chemically inactive and is therefore 
detrimental for sequencing and DNA-based computation, hence a number 
of recent works have looked to avoid them through coding 
\cite{MilKas05,BenBan22,NguCaiKiaDaoSch2323,BarKobLeiYaa23}. Herein we 
focus on relatively long structures, but unlike recent works, we do 
not consider only perfect structures, but also attempt to avoid 
ones which contain impairments, i.e., imperfect structures. We show 
that this problem is closely connected to the above-described channel; 
thus, we are able to also present an efficient encoder for this 
setting.

The paper is organized as follows. In \cref{sec:contr}, we discuss 
the main contribution of this paper, in context of related works. In 
\cref{sec:prelim} we then present necessary notation. Then, in 
\cref{sec:ham} we study the suggested model with substitution 
errors, and in \cref{sec:del} with deletion errors. Finally, in 
\cref{sec:secstruct} we develop an encoder for avoiding the formation 
of even imperfect secondary structures.

\section{Related works and main contribution}\label{sec:contr}

Given a string of length~$n$, the problem of reconstructing it from 
the multiset of (all-, or, in some works, most-) its substrings of a 
fixed length $\ell\leq n$, has been studied in literature. 
Assuming no errors occur in $\bfx$ prior to sampling of its 
substrings, the problem of interest is identifying a set of 
constraints on the information string, equivalent or sufficient, for 
such reconstruction to be achievable.

It was observed in \cite{Ukk92} that under certain circumstances, 
distinct information strings in which repetitions of $\ell$-substrings 
appear in different positions, exhibit the same multisets of 
$(\ell+1)$-substrings. These observations indicate that care must be 
taken when including code-words which contain repeating 
$\ell$-substrings (where observations are made via the multiset of 
$\ell'$-substrings, for some $\ell'\leq \ell+1$). On the other hand, 
if every $\ell$-substring of $\bfx$ is unique, then $\bfx$ is uniquely 
reconstructible from the multiset of its $(\ell+1)$-substrings (and in 
fact, $\ell'$-substrings, for all $\ell'>\ell$), as evident from a 
greedy reconstruction algorithm (which at each stage searches for the 
next/previous character in the information string). This observation 
motivates the study of \emph{repeat-free strings}; $\bfx$ is said to 
be $\ell$-repeat-free if every $\ell$-substring of $\bfx$ is unique 
(put differently, if $\bfx$ is of length~$n$, then it contains 
$n-\ell+1$ distinct $\ell$-substrings).

Focus on repeat-free strings is further justified by the following 
results. It was observed in \cite{ChaChrEzeKia17}, via introduction of 
\emph{profile vectors}, that over an alphabet of size~$q$, where the 
length~$n$ of strings grows, if $\ell<\frac{\log_q(n)}{1+\epsilon}$ 
then the rate of all possible $\ell$-substring multisets vanishes. 
Conversely, it was demonstrated in \cite{EliGabMedYaa21} using 
probabilistic arguments that the asymptotic redundancy of the 
code-book consisting of all $\ell$-repeat-free strings of length~$n$ 
(which, as noted above, is an upper bound for the redundancy of a 
code assuring reconstruction from $(\ell+1)$-substrings), is 
$O(n^{2-\ell/\log_q(n)})$; thus, when $\ell>(1+\epsilon)\log_q(n)$, 
the rate of repeat-free strings alone is~$1$.

In this paper, we extend the setting of previous works by allowing 
information strings to suffer a bounded number of errors, prior to the 
sampling of their substrings. We study this model under two separate 
error models: substitution (Hamming) errors, and deletion errors. In 
both cases we show (see \cref{thm:ham,thm:del}) that when $\ell > 
(1+\epsilon)\log_q(n)$ and the fraction of errors in the substring 
length~$\ell$ is sufficiently small, the rate of generalized 
repeat-free strings, dubbed \emph{resilient-repeat-free}, suffers no 
penalty from the process of sampling, or from the presence of noise 
(when compared to the results of \cite{EliGabMedYaa21}); i.e., the 
required added redundancy is sub-linear. In the case of Hamming noise, 
we also show that when the fraction of errors is too large, 
resilient-repeat-free strings do not exist. However, it is left for 
future works to determine the precise transition between the two 
regimes. Further, we develop an efficient encoder for 
resilient-repeat-free sequences (see \cref{alg:rrfs-encode}), although 
our encoder does not output sequences of a fixed length~$n$, but 
rather only guarantees that the output is of length \emph{at 
most}~$n$.

It should be noted that \cite{GabMil19} presented almost explicit 
encoding/decoding algorithms for codes with a similar noise model. 
However, in that paper's setting, substitution noise affects 
individual substrings \emph{after} sampling; the codes it constructs 
are capable of correcting a constant number of errors in each 
substring, but requires the assumption that errors do not affect the 
same information symbol in a majority of the substrings that reflect 
it. Therefore, its setting is incompatible with the one considered 
herein, whereby each error occurring \emph{before} sampling affects 
$\ell$ consecutive substrings. 
\cite{MarYaa21} also developed codes with full rate, capable of 
correcting a fixed number of errors, occurring in substrings 
independently after sampling. It replaced the aforementioned 
restriction by a constraint on the number of total erroneous 
substrings, which is at most logarithmic in the information string's 
length. Hence, the total number of errors in its setting remains 
asymptotically smaller than the one incurred in the setting considered 
here.

Finally, as mentioned above we exploit the similarity between the 
aforementioned setting and channel model and the problem of avoiding 
secondary structures. We focus on hairpin-loop structures with long 
stems (scaling logarithmically in the length of the sequence), and 
unlike recent works \cite{MilKas05,BenBan22,NguCaiKiaDaoSch2323,
BarKobLeiYaa23} the encoder we develop prevents the formation of such 
structures even when the underlying complementary-base-pair 
hybridization (in a region called the \emph{stem} of the structure) is 
imperfect, that is, it contains at most a $\delta$-fraction of 
mismatched nucleobases (which cannot stably hybridize), while 
asymptotically achieving full rate.

\section{Preliminaries}\label{sec:prelim}

Let $\Sigma^*$ be the set of finite strings over an alphabet $\Sigma$, 
which for convenience we assume to be a finite unital ring of size~$q$ 
(e.g., $\Z_q$). For $\bfx = x(0)x(1)\cdots x(n-1)\in\Sigma^*$, we let 
$\abs*{\bfx}=n$ denote the \emph{length} of $\bfx$. We note that 
indices in the sequel are numbered $0,1,\ldots$. For $\bfx,\bfy\in 
\Sigma^*$, we let $\bfx \bfy$ be their concatenation. For $I\subseteq 
\N$ (we follow the convention $0\in \N$) and $\bfx\in \Sigma^*$, we 
denote by $\bfx_I$ the restriction of~$\bfx$ to indices in~$I$ 
(excluding any indices $\abs*{\bfx}\leq i\in I$), ordered according to 
the naturally inherited order on $I$.

We let $\abs*{A}$ denote the size of a finite set $A$. For a code 
$C\subseteq\Sigma^n$, we define its \emph{redundancy} $\red(C) \deq 
n-\log_q\abs*{C}$, and \emph{rate} $R(C)\deq \frac{1}{n}\log_q\abs*{C} 
= 1-\frac{\red(C)}{n}$.

For $n\in\N$, denote $[n]\deq \bracenv*{0,1,\ldots,n-1}$. Although 
perhaps confusable, for $m\leq n\in\N$ we use the common notation 
$[m,n]\deq \bracenv*{m,m+1,\ldots,n}$. We shall interpret $\bfx_I$ as 
enumerated by $[\abs{I}]$, i.e., $x_I(0) = x(\min I)$, etc. 
Where it is convenient, we will also assume $I\subseteq\N$ to be 
enumerated by $[\abs{I}]$, such that the order of elements is 
preserved; i.e., $I=\mathset*{I(i)}{i\in[\abs{I}]}$, and for all $i\in 
[\abs{I}-1]$ one has $I(i)<I(i+1)$. Under this convention we have, 
e.g., $x_I(0) = x(I(0))$. 
We follow the standard group notation in denoting for $j\in\N$ and 
$I\subseteq\N$, the \emph{coset} $j+I \deq \mathset*{j+i}{i\in I}$.

\begin{example}\label{exm:substring}
Consider the string $\bfx = 0000111101100101$ of length~$n=16$, and 
the set $I = [7,10] = \bracenv*{7,8,9,10} = 7 + [4]$. Then, $\bfx_I = 
1011$, and in particular $\bfx_I(1) = \bfx(I(1)) = \bfx(8) = 0$.
\end{example}

For $\bfx\in\Sigma^*$ and $i,\ell\in\N$, where $i+\ell\leq 
\abs*{\bfx}$, we say that $\bfx_{i+[\ell]}$ is the length-$\ell$ 
\emph{substring} of~$\bfx$ at index~$i$, or \emph{$\ell$-mer} (at 
index~$i$) for short. 
Using notation from \cite{Ukk92}, for $\bfx\in\Sigma^*$ and 
$\ell\in\N$ we denote the multiset of $\ell$-mers of~$\bfx$ by 
\[
Z_\ell(\bfx)\deq 
\multset*{\bfx_{i+[\ell]}}{0\leq i\leq\abs*{\bfx}-\ell}.
\]
We follow \cite{EliGabMedYaa21} in denoting the set of 
\emph{$\ell$-repeat-free} strings 
\[
\rf_\ell(n) \deq \mathset*{\bfx\in\Sigma^n}{i<j \implies 
\bfx_{i+[\ell]}\neq \bfx_{j+[\ell]}}.
\]

We can now more formally state the objectives of 
\cref{sec:ham,sec:del}. 
Assuming an underlying error model, known in context but yet to be 
determined, we let $\ball_t(\bfx)$, for some $\bfx\in\Sigma^*$, be the 
set of strings $\bfy\in\Sigma^*$ which may be the product of at 
most~$t$ errors occurring to $\bfx$. 
Using this notation, our aim shall be to study and design codes 
$C\subseteq\Sigma^n$, such that given $\bfx\in C$ and $\bfy\in 
\ball_t(\bfx)$, for some fixed (or bounded)~$t$, $\bfx$ can be 
uniquely reconstructed given only $Z_\ell(\bfy)$. 
We shall study constraints, generalizing the notion of repeat-free 
strings, which allow unique reconstruction of~$\bfy$, ascertain their 
required redundancy utilizing a probabilistic method, devise explicit 
encoding/decoding algorithms when possible, and state in 
\cref{cor:encoding} specific cases where this in turn allows 
reconstruction of~$\bfx$.

Our analysis of the number of constrained sequences is aided in both 
the Hamming-errors and deletions cases by the following notation:

\begin{definition}
For positive $\ell\leq n$, denote $\binom{[n]}{\ell}\subseteq 2^{[n]}$ 
the collection of $\ell$-subsets of $[n]$. A pair of subsets $(I,J)
\in\binom{[n]}{\ell}^2$ is said to be \emph{observable} if $I(k) < 
J(k)$ for all $k\in [\ell]$.
\end{definition}

Given a string $\bfx\in\Sigma^n$, known from context, we will denote 
for an observable pair $(I,J)\in\binom{[n]}{\ell}^2$ 
\begin{align}\label{eq:u}
\bfu_{I,J} \deq \bfx_I-\bfx_J \in \Sigma^\ell.
\end{align}
We also denote 
$
\Gamma_I \deq 
\lbrace(P,Q) : (P,Q)\;\text{is observable}, \linebreak
(P\cup Q)\cap I \neq \emptyset\rbrace
$.
To simplify notation, where some $\ell\leq n$ is also given, we shall 
abbreviate $\bfu_{i,j}\deq \bfu_{i+[\ell],j+[\ell]}$ and 
\begin{align}\label{eq:gammai}
\Gamma_i\deq \mathset*{(p,q)}{\min(\abs*{i-p}, \abs*{i-q}) < \ell}, 
\end{align}
for any $0\leq i<j\leq n-\ell$.

Then the following lemma will prove useful when bounding the 
redundancy of constrained sequences:

\begin{lemma}\label{lem:ind}
Take $\ell\leq n$ and an observable pair $(I,J)\in 
\binom{[n]}{\ell}^2$. 
Further, let $\bfx\in\Sigma^n$ be chosen uniformly at random. Then 
$\bfu_{I,J}$ is distributed uniformly and mutually independent of 
$
\mathset*{\bfu_{P,Q}}{(P,Q)\not\in \Gamma}_I
$.
\end{lemma}
\begin{IEEEproof}
First, since $\bfu_{I,J}$ is the image of~$\bfx$ under a linear map 
(more precisely, a module homomorphism), the pre-image of any point is 
a coset of the map's kernel and, thus, of equal size; as a result, 
$\bfu_{I,J}$ is distributed uniformly on the map's range. 
Since $(I,J)$ is observable, the map is surjective onto $\Sigma^\ell$, 
hence the first part is completed.

Second, observe that $\bfx_I$ is independent of 
$\bfx_{[n]\setminus I}$, hence mutually independent of 
$\mathset*{\bfu_{P,Q}}{(P,Q)\not\in \Gamma_I}$. 
Since given $\bfx_{[n]\setminus I}$, there exist a bijection between 
$\bfx_I$ and $\bfu_{I,J}$, the proof is concluded.
\end{IEEEproof}

Finally, our proof strategy for bounding the redundancy of said 
constraints is based on Lov\'{a}sz's local lemma (LLL), which we 
slightly rephrase below.
\begin{theorem}\cite[Th.~1.1]{Spe77}\label{thm:LLL}
Let $\bracenv*{A_{i,j}}_{i,j}$ be events in a probability 
space~$\Omega$. If for all $i,j$ there exist constants $0<f_{i,j}<1$ 
such that 
\begin{align*}
\Pr\parenv*{A_{i,j}} 
\leq f_{i,j} \prod_{\mathclap{(p,q)\in \Gamma_{i,j}}} (1-f_{p,q}),
\end{align*}
where $\Gamma_{i,j}$ is such that the event $A_{i,j}$ is mutually 
independent of events $\mathset{A_{p,q}}{(p,q)\not\in \Gamma_{i,j}}$, 
then 
\begin{align*}
\Pr\parenv*{\Omega\setminus\bigcup\big._{i,j} A_{i,j}} 
\geq \prod_{i,j} (1-f_{i,j}).
\end{align*}
\end{theorem}

To the best of authors' knowledge, this application of the lemma is 
novel to the conference version of this work; it then also appeared in 
similar form in a concurrent journal version of~\cite{EliGabMedYaa21}. 
Before continuing, we derive a corollary of \cref{thm:LLL} which is 
less tight, but more easily utilized.

\begin{corollary}\label{cor:LLL}
Let $\bracenv*{A_{i,j}}_{i,j}$ be events in a probability 
space~$\Omega$. If for all $i,j$ there exist constants $0 < \phi_{i,j} 
< 1$ such that 
\begin{align*}
\Pr\parenv*{A_{i,j}} 
\leq \phi_{i,j} \exp\parenv[\Big]{-\sum_{\mathclap{(p,q)\in 
\Gamma_i}} \phi_{p,q} - \phi_{i,j}},
\end{align*}
where $\Gamma_i$ is such that the event $A_{i,j}$ is mutually 
independent of events $\mathset{A_{p,q}}{(p,q)\not\in \Gamma_i}$, then 
\begin{align*}
\Pr\parenv*{\Omega\setminus\bigcup\big._{i,j} A_{i,j}} 
\geq \exp\parenv[\Big]{-\sum_{i,j} \phi_{i,j}}.
\end{align*}
\end{corollary}
\begin{IEEEproof}
The proof follows from the inequality $1-f\geq e^{-f/(1-f)}$ for 
$0<f<1$. Then, denoting $f_{i,j}\deq \frac{\phi_{i,j}}{1+\phi_{i,j}}$ 
and $\Gamma_{i,j}\deq \Gamma_i$ for all $i,j$ we have, for all $i,j$:
\begin{align*}
f_{i,j} \prod_{\mathclap{(p,q)\in \Gamma_{i,j}}} (1-f_{p,q}) 
&= \phi_{i,j} (1-f_{i,j}) 
\prod_{\mathclap{(p,q)\in \Gamma_{i,j}}} (1-f_{p,q}) \\
&\geq \phi_{i,j} \exp\parenv[\Big]{-\phi_{i,j} - \sum_{\mathclap{(p,q)
\in \Gamma_i}} \phi_{p,q}} 
\geq \Pr\parenv*{A_{i,j}}.
\end{align*}
It then follows from \cref{thm:LLL} that 
\begin{IEEEeqnarray*}{+rCl+x*}
\Pr\parenv*{\Omega\setminus\bigcup\big._{i,j} A_{i,j}} 
&\geq& \prod_{i,j} \parenv*{1-\frac{\phi_{i,j}}{1+\phi_{i,j}}} \\
&=& \prod_{i,j} \frac{1}{1+\phi_{i,j}}, 
\end{IEEEeqnarray*}
which, together with $1+\phi\leq e^\phi$ for all $\phi$, concludes the 
proof.
\end{IEEEproof}

Our aim in the next two sections will be to give a precise definition 
to the resilient-repeat-free constraint in the contexts of 
Hamming-errors and deletions respectively, apply \cref{cor:LLL} to 
bound their redundancies, then study explicit encoders (in 
\cref{sec:ham}) and the cases in which error-correcting codes can be 
embedded in these constraints.

\section{Substitution noise}\label{sec:ham}

In this section we consider substitution noise, with error balls 
$\ball[s]_t(\bfx)\deq \mathset*{\bfy}{\ds(\bfx,\bfy)\leq t}$, where 
$\ds(\bfx,\bfy)$ denotes the Hamming distance between $\bfx$ and 
$\bfy$. Observe that the superscript $\mathrm{s}$ denotes 
\emph{substitution} noise, and is not a parameter in this notation.

We present and study a family of repeat-free strings which are 
resilient to substitution errors:
\begin{definition}\label{def:rrf-sub}
We say that $\bfx\in\Sigma^*$ is \emph{$(t,\ell)$-resilient repeat 
free} if the result of any~$t$ substitution errors to $\bfx$ is 
$\ell$-repeat-free. More precisely, we define 
\begin{align*}
\rrfs_{t,\ell}(n) \deq \mathset*{\bfx\in\Sigma^n}{\ball[s]_t(\bfx) 
\subseteq \rf_\ell(n)}.
\end{align*}
Throughout the paper, we shall abbreviate our notation to $\rrfs(n)$, 
given that $t,\ell$ are known from context.
\end{definition}

\begin{example}\label{exm:resilient-rf}
The sequence $\bfx = 0000111101100101$ from \cref{exm:substring} is 
$4$-repeat-free, since all of its substrings of length~$4$ are unique. 
It is not, however, $(1,4)$-resilient-repeat-free, since after a 
single substitution one may derive $\bfy = 
000011110\underline{0}100101$, and $\bfy_{7+[4]} = 1001 = 
\bfy_{10+[4]}$.
\end{example}

\subsection{Rate of resilient-repeat-free strings}

In the following section we dedicate ourselves to study 
$\red(\rrfs(n))$, where $t,\ell$ are taken to be functions of~$n$. In 
particular, we will be interested in developing sufficient (and to a 
lesser degree, necessary) conditions on~$t,\ell$ that assure 
$R(\rrfs(n)) = 1 - o_n(1)$.

Recall that \cite{EliGabMedYaa21} showed that if $\ell = a\log(n) + 
o(\log(n))$, then 
\begin{align*}
R(\rf_\ell(n)) = \begin{cases}
o_n(1), & a<1; \\
1-o_n(1), & a>1.
\end{cases}
\end{align*}
Since $\rrfs_{t,\ell}(n)\subseteq \rrfs_{0,\ell}(n) = \rf_\ell(n)$, 
then with the above scaling of~$\ell$, having $a<1$ implies that 
$R(\rrfs_{t,\ell}(n)) = o_n(1)$ as well, for all~$t$; we shall see 
that when $a>1$, then for sufficiently small~$t$ we still have 
$R(\rrfs_{t,\ell}(n)) = 1 - o_n(1)$.

A particular notion that will aid in our analysis is the following: 
for $0<k\leq \ell$, denote 
\begin{align*}
\cA^\ell_t(k) &\deq \mathset*{\bfx\in \Sigma^{\ell+k}}{\exists\bfy\in 
\ball[s]_t(\bfx): \bfy_{[\ell]} = \bfy_{k+[\ell]}}.
\end{align*}
We let $\pi^\ell_t(k) \deq q^{-(\ell+k)} \abs*{\cA^\ell_t(k)}$ (i.e., 
$\pi^\ell_t(k) = \Pr\parenv*{\bfx\in \cA^\ell_t(k)}$ where $\bfx\in 
\Sigma^{\ell+k}$ is chosen uniformly at random). 
This notion captures the pertinent range of~$k$, since as $k\geq \ell$ 
grows, $\pi^\ell_t(k)$ is clearly fixed and no longer changes 
with~$k$. 
For convenience, when~$\ell,t$ are known from context, we also 
abbreviate:
\begin{align}\label{eq:def-pi}
\pi\deq \pi^\ell_t(\ell);\qquad
\pi'\deq \max_{0<k<\ell} \pi^\ell_t(k).
\end{align}

The usefulness of the notation in~\eqref{eq:def-pi} is substantiated 
in the following theorem.
\begin{theorem}\label{thm:ham-red}
Let $\ell = \ell(n), t = t(n)$ be integer functions, and assume $t\leq 
\ell\leq n$. If for all sufficiently large~$n$ it holds that $3\ell^2 
\pi' + 2\ell n \pi \leq 1/e$, then 
\begin{align*}
\red\parenv*{\rrfs(n)} = O\parenv*{n\log(n) \pi' + n^2 \pi}.
\end{align*}
\end{theorem}
\begin{IEEEproof}
As mentioned above, we shall rely on \cref{cor:LLL}, for which we need 
to define the sets $\bracenv*{A_{i,j}}$, determine the constants 
$\bracenv*{\phi_{i,j}}$ and establish the independence property for 
the sets $\bracenv*{\Gamma_i}$. 
We define for all $0\leq i<j\leq n-\ell$ the sets 
\begin{align*}
A_{i,j} &\deq \mathset*{\bfx\in\Sigma^n}{\exists\bfy\in 
\ball[s]_t(\bfx): \bfy_{i+[\ell]} = \bfy_{j+[\ell]}}.
\end{align*}
Note that $\Sigma^n\setminus\rrfs(n) = \bigcup_{i,j} A_{i,j}$.

We let $\bfx\in\Sigma^n$ be chosen uniformly at random. Then 
$\Pr\parenv*{\bfx\in A_{i,j}} = \pi^\ell_t(\min\bracenv*{\ell,j-i})$. 
Further, 
\begin{align*}
\abs*{\rrfs(n)} 
= q^n \cdot \Pr\parenv*{\bfx\in\rrfs(n)},
\end{align*}
and hence 
\begin{align*}
\red\parenv*{\rrfs(n)} 
= -\log_q\Pr\parenv*{\bfx\in \rrfs(n)}.
\end{align*}
Note that, in our notation, 
$\Pr\parenv*{\bfx\in\rrfs(n)} = 
\Pr\parenv*{\bfx\not\in\bigcup\big._{i,j} A_{i,j}}$.

Recalling \eqref{eq:gammai}, we claim for $0\leq i<j\leq n-\ell$ that 
the event $\bracenv*{\bfx\in A_{i,j}}$ is mutually independent of the 
events $\mathset{\bracenv*{\bfx\in A_{p,q}}}{(p,q)\not\in \Gamma_i}$. 
Indeed, \cref{lem:ind} then implies that $\bfu_{i,j}$ is mutually 
independent of $\mathset{\bfu_{p,q}}{(p,q)\not\in \Gamma_i}$. 
By abuse of notation, consider the mapping $U_{i,j}\colon \bfx\mapsto 
\bfu_{i,j}$; then $\bfx\in A_{i,j}$ if and only if $\bfu_{i,j}\in 
U_{i,j} \ball[s]_t\parenv[\big]{U_{i,j}^{-1}\bfzero}$, where 
$U_{i,j}^{-1}\bfzero = \mathset*{\bfy}{U_{i,j}\bfy = \bfzero}$, 
$\ball[s]_t(A) = \bigcup_{\bfy\in A} \ball[s]_t(\bfy)$, and $U_{i,j}A 
= \mathset*{U_{i,j}\bfy}{\bfy\in A}$. 
Since the sets $U_{i,j} \ball[s]_t\parenv[\big]{U_{i,j}^{-1}\bfzero}$ 
depend only on $t,i,j$ but not $\bfx$, the independence property 
holds. 

Observe that the number of pairs~$(p,q)\in \Gamma_i$ satisfying 
$\abs*{p-q}<\ell$ is over-counted as all choices of $\alpha\in 
[i-\ell+1, i+\ell-1]$ and $\beta\in [\alpha-\ell+1,\alpha+\ell-1]
\setminus \bracenv*{\alpha}$ (then, $(p,q) = (\min\bracenv*{\alpha,
\beta}, \max\bracenv*{\alpha,\beta})$); in fact, this way one counts 
all pairs in $[i-\ell+1,i]$, and all pairs in $[i,i+\ell-1]$, twice 
(in fact, more pairs are counted twice, but the precise number is 
immaterial). I.e., that number is at most $(2\ell-1)(2\ell-2) - 
2\binom{\ell}{2} = (3\ell-2) (\ell-1) < 3\ell^2$.
The number of pairs $(p,q)\in \Gamma_i$ such that $\abs*{q-p}\geq 
\ell$ can also be counted as above, allowing $\beta\in [n]\setminus 
[\alpha-\ell+1,\alpha+\ell-1]$ (which at worst, when $\alpha\in 
\bracenv*{0,n-1}$, allows for $n-\ell+1$ distinct choices), i.e., it 
is at most $(2\ell-1)(n - \ell+1) < 2 \ell n$.

We shall apply an almost symmetric version of \cref{cor:LLL}, where 
\begin{align*}
\phi_{i,j} = \begin{cases}
e \pi, & j-i \geq \ell, \\
e \pi', & j-i < \ell.
\end{cases}
\end{align*}
Then, for any $i,j$ we observe 
\begin{align*}
\phi_{i,j} \exp\parenv[\Big]{-\sum_{\mathclap{(p,q)\in 
\Gamma_i}} \phi_{p,q} - \phi_{i,j}} 
&> \phi_{i,j} e^{-(3\ell^2 \pi' + 2\ell n \pi) e} \\
&\geq \pi^\ell_t(\min\bracenv*{\ell,j-i}) \\
&= \Pr\parenv*{\bfx\in A_{i,j}},
\end{align*}
where the last inequality is justified by $3\ell^2 \pi' + 2\ell n \pi 
< 1/e$, for large enough~$n$. It follows from \cref{cor:LLL} that 
\begin{align*}
\red\parenv*{\rrfs(n)} 
&= -\log_q\Pr\parenv*{\bfx\in\rrfs(n)} \\
&= -\log_q\Pr\parenv*{\bfx\not\in\bigcup\big._{i,j} A_{i,j}} \\
&\leq \log_q(e) \sum\big._{i,j} \phi_{i,j} \\
&< e \log_q(e) \parenv*{n \ell \pi' + n^2 \pi},
\end{align*}
which concludes the proof.
\end{IEEEproof}

Based on the last theorem, it is of interest to bound $\pi, \pi'$ from 
above. 
Our strategy will be twofold. First, we will devise sufficient 
conditions for $\bfx\in \rrfs(n)$; second, we make tighten the 
resulting bounds by taking advantage of the periodicity implied for 
$\bfy\in \ball[s]_t(\bfx)$, by $\bfy_{[\ell]} = \bfy_{k+[\ell]}$. 
To that end, we note the following result.
\begin{lemma}\label{lem:ham-dist}
Take $t\leq \ell\leq n\in\N$, $\bfx\in\Sigma^n$. If for all $0\leq i < 
j\leq n-\ell$ it holds that 
\begin{align*}
\ds\parenv*{\bfx_{i+[\ell]}, \bfx_{j+[\ell]}} 
> t + \max\bracenv*{0,\min\bracenv*{t, \ell-j+i}},
\end{align*}
then $\bfx\in \rrfs(n)$.
\end{lemma}
\begin{IEEEproof}
The proof follows from applying the triangle inequality by cases on 
$(i+[\ell])\cap(j+[\ell])$. Assume to the contrary that there exist 
$\bfy\in \ball[s]_t(\bfx)$ and $0\leq i<j\leq n-\ell$ such that 
$\bfy_{i+[\ell]} = \bfy_{j+[\ell]}$. Note that
\begin{IEEEeqnarray*}{+rCl+x*}
\ds\parenv*{\bfx_{i+[\ell]}, \bfx_{j+[\ell]}} 
&\leq& \ds\parenv*{\bfx_{i+[\ell]}, \bfy_{i+[\ell]}} \>+ \\*
&& \ds\parenv*{\bfy_{i+[\ell]}, \bfy_{j+[\ell]}} \>+ \\*
&& \ds\parenv*{\bfy_{j+[\ell]}, \bfx_{j+[\ell]}} \\
&=& \ds\parenv*{\bfx_{i+[\ell]}, \bfy_{i+[\ell]}} \>+ \\*
&& \ds\parenv*{\bfy_{j+[\ell]}, \bfx_{j+[\ell]}}.
\end{IEEEeqnarray*}
We continue by cases.

If $j-i\geq \ell$ then, since $(i+[\ell])\cap(j+[\ell])=\emptyset$, 
then $\ds\parenv*{\bfx_{i+[\ell]}, \bfy_{i+[\ell]}} 
+ \ds\parenv*{\bfy_{j+[\ell]}, \bfx_{j+[\ell]}} \leq t$, which 
contradicts the theorem's assumption.

On the other hand, if $j-i\leq \ell-t$, then we may simply bound 
$
\ds\parenv*{\bfx_{i+[\ell]}, \bfy_{i+[\ell]}} 
+ \ds\parenv*{\bfy_{j+[\ell]}, \bfx_{j+[\ell]}} \leq 2t
$,
again in contradiction.

Finally, suppose $\ell-t < j-i < \ell$. Note that 
\begin{IEEEeqnarray*}{+rCl+x*}
\IEEEeqnarraymulticol{3}{l}{\ds\parenv*{\bfx_{i+[\ell]}, 
\bfy_{i+[\ell]}} + \ds\parenv*{\bfy_{j+[\ell]}, \bfx_{j+[\ell]}}} \\
\quad &=& \ds\parenv*{\bfx_{[i,j-1]}, \bfy_{[i,j-1]}} 
+ 2\ds\parenv*{\bfx_{[j,i+\ell-1]}, \bfy_{[j,i+\ell-1]}} \>+ \\*
&& \ds\parenv*{\bfy_{[i+\ell,j+\ell-1]}, \bfx_{[i+\ell, j+\ell-1]}}.
\end{IEEEeqnarray*}
Since $[i,j-1], [j,i+\ell-1], [i+\ell,j+\ell-1]$ are pairwise 
disjoint, 
\begin{IEEEeqnarray*}{+rCl+x*}
\ds\parenv*{\bfx_{[i,j-1]}, \bfy_{[i,j-1]}} &+& 
\ds\parenv*{\bfx_{[j,i+\ell-1]}, \bfy_{[j,i+\ell-1]}} \>+ \\*
&& \> \ds\parenv*{\bfy_{[i+\ell,j+\ell-1]}, \bfx_{[i+\ell,j+\ell-1]}} 
\leq t.
\end{IEEEeqnarray*}
Hence, denoting $\Delta\deq \ds\parenv*{\bfx_{[j,i+\ell-1]}, 
\bfy_{[j,i+\ell-1]}}$, we have 
\begin{align*}
\ds\parenv*{\bfx_{i+[\ell]}, \bfx_{j+[\ell]}} 
&\leq t + \Delta \leq t + (\ell-j+i),
\end{align*}
once more in contradiction. This concludes the proof.
\end{IEEEproof}

Observe in particular that \cref{lem:ham-dist} applies to $n=\ell+k$, 
and its proof can be applied specifically for $(i,j)=(0,k)$. That is, 
if $\wt(\bfu_{0,k}) = \ds(\bfx_{[\ell]}, \bfx_{k+[\ell]}) > t + 
\min\bracenv*{t, \ell-k}$ then $\bfx\not\in \cA^\ell_t(k)$. Vice 
versa, $\pi^\ell_t(k) = \Pr\parenv*{\bfx\in \cA^\ell_t(k)}\leq 
\Pr\parenv*{\wt(\bfu_{0,k})\leq t + \min\bracenv*{t, \ell-k}}$, which 
leads to the following bound:

\begin{corollary}\label{cor:naive-pi-bound}
$\pi^\ell_t(k)\leq q^{-\ell} \sum_{i=0}^{t + \min\bracenv*{t, \ell-k}} 
\binom{\ell}{i} (q-1)^i$.
\end{corollary}
\begin{IEEEproof}
By \cref{lem:ind} $\bfu_{0,k}\in \Sigma^\ell$ is distributed 
uniformly, hence from the above observation the proof is concluded.
\end{IEEEproof}

The bound of \cref{cor:naive-pi-bound} can be improved upon in some 
cases, depending on~$k$ (thus improving the upper bound on~$\pi'$):
\begin{lemma}\label{lem:imp-pi}
\begin{align*}
\pi^\ell_t(k) \geq q^{-\ell} \sum_{i=0}^t \binom{\ell}{i}(q-1)^i,
\end{align*}
and
\begin{align*}
\pi^\ell_t(k) 
\leq \begin{cases}
q^{-\ell} \sum_{i=0}^t \binom{\ell+k}{i}(q-1)^i, & k 
\leq \tfrac{\ell}{2}; \\
q^{-\ell} \sum_{i=0}^t \binom{2\ell-k}{i}(q-1)^i & k > \tfrac{\ell}{2}.
\end{cases}
\end{align*}
\end{lemma}
\begin{IEEEproof}
Take integers $p\geq 2$ and $0\leq r<k$ such that $\ell+k = pk+r$. For 
$\bfx\in \cA^\ell_t(k)$, there exists $\bfy\in \ball[s]_t(\bfx)$ such 
that $\bfy_{[\ell]} = \bfy_{k+[\ell]}$. The method of our proof 
utilizes the observation $\bfy_{[\ell]} = \bfy_{k+[\ell]}$ implies 
that~$\bfy$ is $k$-periodic, i.e., can be determined by its first~$k$ 
coordinates:
\begin{align*}
\bfy = (\underbrace{\bfy_{[k]}, \ldots, \bfy_{[k]}}_{p\ \text{times}}, 
\bfy_{[r]}).
\end{align*}

Observe for each $\bfy\in \Sigma^{\ell+k}$, satisfying $\bfy_{[\ell]} 
= \bfy_{k+[\ell]}$ (of which we have seen there exist precisely $q^k$ 
distinct possibilities, corresponding to a free choice of 
$\bfy_{[k]}$), that one may form a unique~$\bfx\in \cA^\ell_t(k)$ by 
changing at most~$t$ of the symbols $\bfy_{k+[\ell]}$. It follows that 
\begin{align*}
\abs*{\cA^\ell_t(k)} \geq q^k \sum_{i=0}^t \binom{\ell}{i} (q-1)^{i}.
\end{align*}
On the other hand, it is also straightforward that 
\begin{align*}
\abs*{\cA^\ell_t(k)} \leq q^k \sum_{i=0}^t \binom{\ell+k}{i} 
(q-1)^i, 
\end{align*}
by changing up to~$t$ symbols of the whole of~$\bfy$.

When $p=2$ or, equivalently, $k>\frac{\ell}{2}$, we shall improve the 
above bound.
Take $\bfx\in \cA_k$ and $\bfy\in\ball[s]_t(\bfx)$ satisfying 
$\bfy_{[\ell]} = \bfy_{k+[\ell]}$. 
Define the intervals $I_1 \deq [\ell-k]$, $I_2\deq [\ell-k,k-1]$, 
$I_3\deq [k,\ell-1]$, $I_4\deq [\ell,2k-1]$, $I_5\deq [2k,k+\ell-1]$. 
Using this notation, we have $\bfy_{I_2} = \bfy_{I_4}$ and $\bfy_{I_1} 
= \bfy_{I_3} = \bfy_{I_5}$, i.e., 
\begin{align*}
\bfy = \bfy_{I_1} \bfy_{I_2} \bfy_{I_1} \bfy_{I_2} \bfy_{I_1}.
\end{align*}
Consider the string 
$
\bfy' \deq \bfy_{I_1} \bfx_{I_2} \bfy_{I_1} \bfx_{I_2} \bfy_{I_1}
$ 
and note that 
\begin{IEEEeqnarray*}{+rCl+x*}
\ds\parenv*{\bfx, \bfy'} 
&=& \ds\parenv*{\bfx_{I_1\cup I_3\cup I_5}, 
\bfy'_{I_1\cup I_3\cup I_5}} \>+ \\*
&& \ds\parenv*{\bfx_{I_2}, \bfy'_{I_2}} 
+ \ds\parenv*{\bfx_{I_4}, \bfy'_{I_2}} \\
&=& \ds\parenv*{\bfx_{I_1\cup I_3\cup I_5}, 
\bfy_{I_1\cup I_3\cup I_5}} \>+ \\*
&& \ds\parenv*{\bfx_{I_2}, \bfx_{I_2}} 
+ \ds\parenv*{\bfx_{I_4}, \bfx_{I_2}} \\
&=& \ds\parenv*{\bfx_{I_1\cup I_3\cup I_5}, 
\bfy_{I_1\cup I_3\cup I_5}} \>+ \\*
&& 0 + \ds\parenv*{\bfx_{I_4}, \bfx_{I_2}}.
\end{IEEEeqnarray*}
Applying the triangle inequality on the last addend, 
\begin{IEEEeqnarray*}{+rCl+x*}
\ds\parenv*{\bfx, \bfy'} 
&\leq& \ds\parenv*{\bfx_{I_1\cup I_3\cup I_5}, 
\bfy_{I_1\cup I_3\cup I_5}} \>+ \\*
&& \ds\parenv*{\bfx_{I_4}, \bfy_{I_4}} 
+ \ds\parenv*{\bfy_{I_4}, \bfx_{I_2}} \\
&=& \ds\parenv*{\bfx_{I_1\cup I_3\cup I_5}, 
\bfy_{I_1\cup I_3\cup I_5}} \>+ \\*
&& \ds\parenv*{\bfx_{I_4}, \bfy_{I_4}} 
+ \ds\parenv*{\bfy_{I_2}, \bfx_{I_2}} \\
&=& \ds(\bfx, \bfy) \leq  t.
\end{IEEEeqnarray*}
Therefore, for any $\bfx\in \cA^\ell_t(k)$ there exists $\bfy'\in 
\ball[s]_t(\bfx)$ of the form 
\begin{align*}
\bfy' = \bfy_{I_1} \bfx_{I_2} \bfy_{I_1} \bfx_{I_2} \bfy_{I_1},
\end{align*}
and in particular $\bfx_{I_2} = \bfy'_{I_2}$. This implies an improved 
upper bound, by freely choosing $\bfy_{I_1} \bfx_{I_2}\in \Sigma^k$ 
and subsequently at most~$t$ coordinates from $[\ell+k]\setminus I_2$ 
to change, as follows
\begin{IEEEeqnarray*}{+rCl+x*}
\abs*{\cA^\ell_t(k)} \leq q^k \sum_{i=0}^t \binom{2\ell-k}{i}(q-1)^i.
\\[-\normalbaselineskip] &&&\IEEEQEDhere
\end{IEEEeqnarray*}
\end{IEEEproof}

With the results of \cref{cor:naive-pi-bound,lem:imp-pi}, we can now 
bound $\pi,\pi'$ to facilitate the application of \cref{thm:ham-red}.

\begin{corollary}\label{cor:imp-pi}
\begin{align*}
\pi = q^{-\ell} \sum_{i=0}^t \binom{\ell}{i}(q-1)^i 
\leq q^{-\ell \parenv*{1 - H_q\parenv*{\min\bracenv*{\frac{q-1}{q}, 
\frac{t}{\ell}}}}},
\end{align*}
and 
\begin{IEEEeqnarray*}{+rCl+x*}
\pi' &\leq& q^{-\ell} \sum_{i=0}^t \binom{\floorenv*{3\ell/2}}{i}
(q-1)^i \\
&\leq& q^{-\ell\parenv*{1 - \frac{3}{2} 
H_q\parenv*{\min\bracenv*{\frac{q-1}{q}, \frac{2t}{3\ell}}}}},
\end{IEEEeqnarray*}
where $H_q(\delta) = \delta \log_q(q-1) - \delta \log_q(\delta) - 
(1-\delta) \log_q(1-\delta)$ is the $q$-ary entropy function.
\end{corollary}
\begin{IEEEproof}
First observe the equality on the first line follows from the lower 
bound of \cref{lem:imp-pi}, together with the upper bound of either 
\cref{cor:naive-pi-bound} or \cref{lem:imp-pi}. Similarly, the first 
inequality on the second line follows from the upper bound of 
\cref{lem:imp-pi}.

The second inequality on both lines follows from the standard bound on 
the size of the $q$-ary Hamming ball (see, e.g., 
\cite[Lem.~4.7]{Rot06}); in particular observe that 
\begin{align*}
\sum_{i=0}^t \binom{\floorenv*{3\ell/2}}{i}(q-1)^i 
\leq q^{\floorenv*{3\ell/2} H_q\parenv*{\min\bracenv*{\frac{q-1}{q}, 
\frac{t}{\floorenv*{3\ell/2}}}}},
\end{align*}
and since $x\mapsto x H_q(1/x)$ is increasing for $x\geq 1$, the claim 
follows.
\end{IEEEproof}
We note before continuing that applying the upper bound of 
\cref{cor:naive-pi-bound} instead of \cref{lem:imp-pi} would result in 
an inferior upper bound on~$\pi'$.

Motivated by the discussion at the beginning of this section, we 
fix the values of $t,\ell$ for the reminder of this paper. 
Take~$a>1$ and a real number~$\delta>0$; we let 
\begin{align}\label{eq:tell}
\ell &\deq \floorenv*{a\log_q(n)}; \nonumber\\
t &\deq \floorenv*{\delta \ell} 
= \floorenv*{\delta \floorenv*{a\log_q(n)}}, 
\end{align}
as~$n$ grows.

Inspired by \cref{cor:imp-pi}, we also denote by~$\tildel$ the 
(unique) real number $0 < \tildel < \frac{q-1}{q}$ satisfying 
\[
H_q\parenv*{\tfrac{2}{3} \tildel} = \tfrac{2}{3}.
\]
We observe by substitution that $\tildel > \frac{q-1}{2q}$, and 
provide~$\tildel$ for some small values of~$q$:

\begin{align*}
\begin{array}{c|c|c|c}
q & \tfrac{q-1}{2q} & \tildel & \tfrac{q-1}{q} \\\hline
2 & 0.25 & 0.2609 & 0.5 \\
3 & 0.3333 & 0.3723 & 0.6667 \\
4 & 0.375 & 0.4375 & 0.75 \\
5 & 0.4 & 0.4817 & 0.8 \\
6 & 0.4167 & 0.5141 & 0.8333
\end{array}
\end{align*}

Applying the result of \cref{cor:imp-pi} to \cref{thm:ham-red}, we can 
now obtain the following result.
\begin{theorem}\label{thm:ham}
Fix $a>1$, $0<\delta<\tildel$. Then, as $n\to\infty$, 
\begin{align*}
\red\parenv*{\rrfs(n)} = O(n^{2-a\parenv*{1-H_q(\delta)}}).
\end{align*}
\end{theorem}
\begin{IEEEproof}
If $a\leq\parenv*{1 - H_q\parenv*{\delta}}^{-1}$ the proposition 
vacuously holds.

Otherwise, let $\bfx\in \Sigma^{\ell+k}$ be chosen uniformly at 
random. 
Based on \cref{cor:imp-pi} (recalling again that $x\mapsto x H_q(1/x)$ 
is increasing for $x\geq 1$), we observe for $\delta < \tildel$ that 
\begin{align*}
\pi &\leq q\cdot n^{-a \parenv*{1 - H_q(\delta)}}; \\
\pi' &\leq q\cdot 
n^{-a \parenv*{1 - \frac{3}{2} H_q(\frac{2}{3}\delta)}}.
\end{align*}
Hence, for sufficiently large~$n$ it holds that $3\ell^2 \pi' + 2 \ell 
n \pi < 1/e$, satisfying the conditions of \cref{thm:ham-red}. Since 
we also have $n\log(n)\pi'=o\parenv*{n^{2-a\parenv*{1-H_q(\delta)}}}$, 
the claim follows from \cref{thm:ham-red}.
\end{IEEEproof}

\begin{corollary}\label{cor:ham-rate}
Take $0<\delta<\tildel$. If $a > \parenv*{1-H_q\parenv*{\delta}}^{-1}$ 
then $R\parenv*{\rrfs(n)} = 1-o(1)$, and if $a \geq 2 \parenv*{1-
H_q(\delta)}^{-1}$, then $\rrfs(n)$ incurs a constant number of 
redundant symbols. 
\end{corollary}

The last corollary can be viewed in the context of related works; as 
mentioned above, \cite{EliGabMedYaa21} demonstrated that if $a>1$ then 
$R\parenv*{\rf_\ell(n)} = 1-o_n(1)$, and if $a\geq 2$ then 
$\red\parenv*{\rf_\ell(n)} = O_n(1)$. 
\cref{cor:ham-rate} demonstrates that if $a>1$ (respectively $a\geq 
2$), then for all sufficiently small $\delta>0$ it holds that 
$R\parenv*{\rrfs(n)} = 1-o_n(1)$ (respectively, 
$\red\parenv*{\rrfs(n)} = O_n(1)$). That is, resilient-repeat-free 
sequences for a number of substitutions errors logarithmic in the 
string length (linear in the substring length) incur no additional 
asymptotic cost.

Up until here, we have focused on demonstrating conditions sufficient 
for the rate of resilient-repeat-free strings to be asymptotically 
optimal. In the sequel, we pursue the converse, or more precisely, 
necessary conditions for such strings to obtain non-vanishing rate.

\begin{definition}\label{def:err-cor-code}
For a real $\delta$, $0\leq\delta<1$, and an integer $\ell>0$, let 
$M_q(\ell,\delta)$ be the maximum number of code-words in a code 
$C\subseteq\Sigma^\ell$ such that $\ds(\bfx, \bfy)\geq \delta \ell$ 
for any distinct $\bfx, \bfy\in C$. For a given $\delta>0$, define the 
maximum achievable rate by
\begin{align*}
R_q(\delta)\deq 
\limsup_{\ell\to\infty} \tfrac{1}{\ell}\log_q M_q(\ell,\delta).
\end{align*}
For completeness, we state the well-known Gilbert-Varshamov and 
Elias-Bassalygo bounds (see, e.g., \cite[Thm.4.9-12]{Rot06}) for 
$\delta \leq \frac{q-1}{q}$, 
\begin{align*}
1-H_q(\delta) \leq R_q(\delta)\leq 
1-H_q\parenv*{\tfrac{q-1}{q}\parenv*{1-\sqrt{1-\tfrac{q}{q-1}
\delta}}}.
\end{align*}
\end{definition}

The following lemma states a converse bound to \cref{cor:ham-rate}.
\begin{lemma}\label{lem:ham-conv}
If $a < R_q(\delta)^{-1}$, then for sufficiently large $n\in\N$ 
\begin{align*}
\rrfs(n) = \emptyset.
\end{align*}
In particular, the statement holds if $t\geq \frac{q-1}{q} \ell$, for 
all~$a$.
\end{lemma}
\begin{IEEEproof}
Take, on the contrary, some $\bfx\in \rrfs(n)$. By \cref{def:rrf-sub}, 
the $\ell$-mers 
\begin{align*}
\mathset*{\bfx_{i \ell + [\ell]}}{0\leq i\leq \floorenv*{n/\ell}-1} 
\subseteq \Sigma^\ell
\end{align*}
form a code of size $\floorenv*{n/\ell}$ and minimum distance $d > t 
\geq \delta \ell$. By \cref{def:err-cor-code} we obtain
\begin{align*}
\frac{\log \floorenv{n/\ell}}{\ell} \leq {R}_q(\delta) + o(1).
\end{align*}
Recalling $\ell=\floorenv*{a \log n}$ yields that
\begin{align*}
\tfrac{1}{a} \leq {R}_q(\delta) + o(1), 
\end{align*}
in contradiction to the assumption.
\end{IEEEproof}

It should be noted that \cref{lem:ham-conv} specifically pertains to 
resilient-repeat-free strings, which the reader will observe are not 
necessarily required for successful reconstruction of information. 
Nevertheless, it might be conjectured, based on the noiseless case, 
that resilient-repeat-free sequences may achieve optimum asymptotic 
rate.

Before concluding, we note that a twofold gap remains between 
\cref{thm:ham} and the converse of \cref{lem:ham-conv}. First, 
$\red\parenv*{\rrfs(n)}$ is not characterized when $R_q(\delta)^{-1} 
\leq a \leq \parenv*{1-H_q\parenv*{\delta}}^{-1}$; and second, it is 
not found when $\delta\geq\tildel$.

\subsection{Encoding resilient-repeat-free codes}\label{sec:rrf-enc}

In this section, we present an explicit encoder of 
resilient-repeat-free strings, in the hope that it may then be 
utilized in constructing error-correcting codes for the noisy 
substring channel.

We first discuss how elements of the ball $\ball[s]_t(\bfzero)$ 
(which throughout this discussion we assume to contain length-$\ell$ 
sequences; observe that we opt to use~$\ell$ instead of~$n$ here 
since our analysis will later be applied to $\ell$-substring of a 
longer length-$n$ string) may be enumerated. Observe that given any 
$\bfx_{[k]}\in \Sigma^k$ with $\wt(\bfx_{[k]})\leq t$, 
\begin{align*}
n(\bfx_{[k]}) &\deq 
\abs*{\mathset*{\bfy\in \ball[s]_t(\bfzero)}{\bfy_{[k]} = 
\bfx_{[k]}}} \\
&= \sum_{j=0}^{t-\wt(\bfx_{[k]})} \binom{\ell-k}{j} (q-1)^j.
\end{align*}

\begin{example}\label{exm:lexi-num}
We take $q = 2, \ell = 7, t = 3, k = 4$ and $\bfx = 0110010\in 
\Sigma^7\cap \ball[s]_3(\bfzero)$. Then, the number of elements 
$\bfy\in \ball[s]_3(\bfzero)$ such that $\bfy_{[4]} = \bfx_{[4]} = 
0110$ equals 
\begin{align*}
n(\bfx_{[4]}) &= 
\sum_{j=0}^{3-\wt(0110)} \binom{7-4}{j} (2-1)^j \\
&= \sum_{j=0}^1 \binom{3}{j}  = 1 + 3 = 4.
\end{align*}
These elements are 
\begin{align*}
\bracenv*{0110000, 0110001, 0110010, 0110100}.
\end{align*}
\end{example}

Assuming a total order~$<$ on~$\Sigma$, denote $\norm{x}\deq 
\abs[\big]{\mathset{y\in \Sigma}{y<x}}$ for all~$x\in \Sigma$. It was 
shown in \cite{Cov73} that the lexicographic index of~$\bfx\in 
\ball[s]_t(\bfzero)$ equals 
\begin{align}\label{eq:lexi-ind}
i(\bfx) 
&= \sum_{k\in [\ell]} \sum_{\alpha < x(k)} n(\bfx_{[k-1]} \alpha), 
\end{align}
where we let $\bfx_{[0]}$ be the empty string, with $\wt(\bfx_{[0]})
\deq 0$.

\begin{example}\label{exm:lexi-ind}
We use the natural order $0<1$ with $q=2$. Then, using $\bfx = 
0110010$ from \cref{exm:lexi-num} we ascertain its lexicographic index 
in $\ball[s]_3(\bfzero)$:
\begin{IEEEeqnarray*}{+rCl+x*}
i(\bfx) 
&=& \sum_{k=0}^6 \sum_{\alpha < x(k)} n(\bfx_{[k]} \alpha) \\
&=& n(\bfx_{[1]} 0) + n(\bfx_{[2]} 0) + n(\bfx_{[5]} 0) \\
&=& n(00) + n(010) + n(011000) \\
&=& \sum_{j=0}^{3-\wt(00)} \binom{7-2}{j} + 
\sum_{j=0}^{3-\wt(010)} \binom{7-3}{j} \>+ \\*
&& \sum_{j=0}^{3-\wt(011000)} \binom{7-6}{j} \\
&=& \sum_{j=0}^3 \binom{5}{j} + 
\sum_{j=0}^2 \binom{4}{j} + 
\sum_{j=0}^1 \binom{1}{j} \\
&=& (1+5+10+10) + (1+4+6) + (1+1) = 39.
\end{IEEEeqnarray*}
\end{example}

\begin{lemma}\label{lem:lexi-ind}
The lexicographic index of~$\bfx\in \ball[s]_t(\bfzero)$ equals 
\begin{IEEEeqnarray*}{+rCl+x*}
i(\bfx) 
&=& \sum_{k\in [\ell]} \norm{x(k)} \cdot n(\bfx_{[k+1]}) \>+ \\*
&& \hphantom{\sum_{k\in [\ell]}} \wt(x(k)) \cdot 
\binom{\ell-k-1}{t-\wt(\bfx_{[k]})} (q-1)^{t-\wt(\bfx_{[k]})}.
\end{IEEEeqnarray*}
\end{lemma}
\begin{IEEEproof}
Observe that if $x(k)\neq 0$, then 
\begin{IEEEeqnarray*}{+rCl+x*}
n(\bfx_{[k]} 0) 
&=& \sum_{j=0}^{t-\wt(\bfx_{[k]} 0)} \binom{\ell-(k+1)}{j} (q-1)^j \\
&=& \sum_{j=0}^{t-\wt(\bfx_{[k]})} \binom{\ell - k - 1}{j} (q-1)^j, 
\end{IEEEeqnarray*}
hence
\begin{IEEEeqnarray*}{+rCl+x*}
n(\bfx_{[k+1]}) 
&=& \sum_{j=0}^{t-\wt(\bfx_{[k+1]})} \binom{\ell-(k+1)}{j} (q-1)^j \\
&=& \sum_{j=0}^{t-\wt(\bfx_{[k]})-1} \binom{\ell - k - 1}{j} (q-1)^j \\
&=& n(\bfx_{[k]} 0) - \binom{\ell - k - 1}{t-\wt(\bfx_{[k]})} 
(q-1)^{t-\wt(\bfx_{[k]})}.
\end{IEEEeqnarray*}
It is also immediate that for any $\alpha\in \Sigma\setminus 
\bracenv*{0}$ it holds that $n(\bfx_{[k]} \alpha) = n(\bfx_{[k+1]})$. 
The claim now follows from \eqref{eq:lexi-ind}.
\end{IEEEproof}

\begin{example}\label{exm:lexi-newind}
Repeating \cref{exm:lexi-ind} with \cref{lem:lexi-ind} we can find 
\begin{IEEEeqnarray*}{+rCl+x*}
i(\bfx) 
&=& \sum_{k\in [7]} \norm{x(k)} \cdot n(\bfx_{[k+1]}) \>+ \\*
&& \hphantom{\sum_{k\in [7]}} \wt(x(k)) \cdot 
\binom{6-k}{t-\wt(\bfx_{[k]})} (q-1)^{3-\wt(\bfx_{[k]})} \\
&=& n(\bfx_{[2]}) + \binom{6-1}{3-\wt(\bfx_{[1]})} \>+ \\*
&& n(\bfx_{[3]}) + \binom{6-2}{3-\wt(\bfx_{[2]})} \>+ \\*
&& n(\bfx_{[6]}) + \binom{6-5}{3-\wt(\bfx_{[5]})} \\
&=& n(01) + \binom{5}{3} \>+ \\*
&& n(011) + \binom{4}{2} \>+ \\*
&& n(011001) + \binom{1}{1} \\
&=& \sum_{j=0}^2 \binom{5}{j} + \binom{5}{3} \>+ \\*
&& \sum_{j=0}^1 \binom{4}{j} + \binom{4}{2} \>+ \\*
&& \binom{1}{0} + \binom{1}{1} = 39,
\end{IEEEeqnarray*}
matching the result of \cref{exm:lexi-ind}.
\end{example}

Computationally, the most taxing expression to calculate in the sum of 
\cref{lem:lexi-ind} is $n(\bfx_{[k+1]})$; however, one might employ a 
recursive approach to obtaining the sum. Indeed, from the Pascal 
identity we observe 
\begin{IEEEeqnarray*}{+rCl+x*}
\IEEEeqnarraymulticol{3}{l}{n(\bfx_{[k]})} \\
&=&  \sum_{j=0}^{t-\wt(\bfx_{[k]})} 
\binom{\ell-k}{j} (q-1)^j \\
&=& \sum_{j=1}^{t-\wt(\bfx_{[k]})} \binom{\ell-k-1}{j-1} (q-1)^j \>+ \\*
&& \sum_{j=0}^{t-\wt(\bfx_{[k]})} \binom{\ell-k-1}{j} (q-1)^j \\
&=& (q-1) \sum_{j'=0}^{t-\wt(\bfx_{[k]})-1} \binom{\ell-k-1}{j'} 
(q-1)^{j'} \>+ \\*
&& \sum_{j=0}^{t-\wt(\bfx_{[k]})} \binom{\ell-k-1}{j} (q-1)^j \\
&=& q \sum_{j=0}^{t-\wt(\bfx_{[k]})-1} \binom{\ell-k-1}{j} (q-1)^j \>+ \\*
&& \binom{\ell-k-1}{t-\wt(\bfx_{[k]})} (q-1)^{t-\wt(\bfx_{[k]})}.
\end{IEEEeqnarray*}
By incrementing the upper limit of the sum through $t-\wt(\bfx_{[k]})$ 
and subtracting the corresponding addend separately, the last line can 
also be restated 
\begin{IEEEeqnarray*}{+rCl+x*}
n(\bfx_{[k]}) &=& 
q \sum_{j=0}^{t-\wt(\bfx_{[k]})} \binom{\ell-k-1}{j} (q-1)^j \>- \\*
&& (q-1)\binom{\ell-k-1}{t-\wt(\bfx_{[k]})} (q-1)^{t-\wt(\bfx_{[k]})}.
\end{IEEEeqnarray*}
Partitioning into cases by $\wt(x(k))$, we find 
\begin{align*}
n(&\bfx_{[k]}) \\
=&\> q\cdot n(\bfx_{[k+1]}) \>- \\*
& (-1)^{\wt(x(k))} \binom{\ell-k-1}{t-\wt(\bfx_{[k]})} 
(q-1)^{t-\wt(\bfx_{[k+1]})+1}, \numberthis\label{eq:lexi-num-rec}
\end{align*}
where trivially $n(\bfx_{[\ell]}) = n(\bfx) = 1$.

\begin{example}
In \cref{exm:lexi-num,exm:lexi-newind} we have found, in 
$\ball[s]_3(\bfzero)$, 
\begin{align*}
n(01) &= \sum_{j=0}^2 \binom{5}{j} = 16; \\
n(011) &= \sum_{j=0}^1 \binom{4}{j} = 5; \\
n(0110) &= \sum_{j=0}^1 \binom{3}{j} = 4.
\end{align*}
We can now confirm that indeed 
\begin{align*}
n(011) &= 2\cdot n(0110) - (-1)^{\wt(0)} \binom{6-3}{3-2} \\
&= 2\cdot 4 - 3, 
\end{align*}
and 
\begin{align*}
n(01) &= 2\cdot n(011) - (-1)^{\wt(1)} \binom{6-2}{3-1} \\
&= 2\cdot 5 + 6.
\end{align*}
\end{example}

It now follows from \cref{lem:lexi-ind} and \eqref{eq:lexi-num-rec} 
that computing the sum for the index~$i(\bfx)$ can be done for $k\in 
[\ell]$ in descending order, where at each addend it is required to:
\begin{enumerate}
\item 
Compute binomial coefficients of the form~$\binom{\ell-k}{j}$ for 
$j\leq t$ (all of order at most~$\binom{\ell}{t}\leq \frac{\ell^t}{t!}
\leq \parenv{\frac{e \ell}{t}}^t$). Observe each binomial coefficient 
requires~$\log\parenv{\frac{e \ell}{t}}^t < t\log(\ell)$ symbols, and 
at most~$t+1$ need to be stored at a time, so that 
$\mathset[\big]{\binom{\ell-k-1}{j}}{j\leq t}$ could be computed via 
the Pascal identity from $\mathset[\big]{\binom{\ell-k}{j}}{j\leq t}$. 
This stage hence requires~$O(t^2 \log(\ell))$ operations, and~$O(t^2 
\log(\ell))$ space.

Further, obtaining $\mathset[\big]{\binom{\ell}{j}}{j\leq t}$ for 
initialization requires at most~$O(t^2 \log(\ell) \ell)$ operations, 
if it is performed similarly.

\item 
Multiplying a binomial coefficient by at most~$q^t$, which may 
practically be performed in~$O(t \log(\ell) \log\log(\ell) \linebreak 
\log\log\log(\ell))$ operations.

\item 
Computing~$n(\bfx_{[k]})$, which has seen above requires $O(t 
\log(\ell))$ space and $O(t \log(\ell) \log\log(\ell) 
\log\log\log(\ell))$ operations.

\item 
Summing the results requires~$O(t \log(\ell))$ operations and~$O(t 
\log(\ell))$ space.
\end{enumerate} 
The entire algorithm therefore requires at most~$O(t^2 \log(\ell) 
\ell)$ operations and~$O(t^2 \log(\ell))$ space. That is, if $t,\ell = 
O(\log(n))$, at most~$O((\log(n))^3 \log\log(n))$ operations 
and~$O((\log(n))^2 \linebreak \log\log(n))$ space.

The inverse operation, obtaining~$\bfx\in \ball[s]_t(\bfzero)$ 
such that $i(\bfx) = i$, for some given~$i$, is also due to 
\cite{Cov73}: starting with the empty sequence for~$k=0$, assume 
$\bfx_{[k-1]}$ has already been constructed for some $0<k\leq \ell$. 
Going over $\alpha\in \Sigma$ in increasing order (assuming the same 
total order as before), if $i\leq n(\bfx_{[k-1]}\alpha)$ then set 
$x(k-1)\deq \alpha$ and update $i\leftarrow i-n(\bfx_{[k-1]}\alpha)$; 
otherwise increase~$\alpha$ and repeat; the maximum element 
of~$\Sigma$ can be filled in without comparison, if the algorithm 
arrives at it. Again, the limiting step of the algorithm is obtaining 
the representation of the binomial coefficients, and while the 
algorithm might require~$t \ell$ steps in the worst case, these do not 
need to be recalculated unless $k$ is increased. Thus, calculating the 
inverse also requires at most~$O((\log(n))^3 \log\log(n))$ operations 
and~$O((\log(n))^2 \log\log(n))$ space, for~$t,\ell = O(\log(n))$.

In summary, we have obtained an explicit and invertible enumerator 
of~$\ball[s]_t(\bfzero)$, with the aforementioned complexity, 
which we denote $\en(\bfx)$. Recall that $\abs*{\ball[s]_t(\boldsymbol 
0)}\leq q^{\ell H_q(t/\ell)}$, i.e., $\en\colon \ball[s]_t(\boldsymbol 
0)\to \Sigma^{\ceilenv*{\ell H_q(t/\ell)}}$.

Equipped with an (efficient) enumeration algorithm 
for~$\ball[s]_t(\bfzero)$, we may now propose an explicit encoder of 
resilient-repeat-free sequences. Our encoder has the drawback that it 
produces sequences in $\bigcup_{m\leq n} \rrfs_{t,\ell}(m)$ (here, $t,
\ell$ are specified to stress that they are invariant in~$m$, 
depending only on~$n$) rather than solely in~$\rrfs(n)$; however, in 
practice this does not seem too onerous for applications, were shorter 
sequences may be stored just as easily, as long as data is 
recoverable.

\begin{algorithm}[t]
\caption{Resilient-repeat-free Encoder}
\label{alg:rrfs-encode}
\SetKwInput{KwInput}{Input}                
\SetKwInput{KwOutput}{Output}  
\SetAlgoLined
\KwInput{ $\bfx\in \Sigma^n$ containing no $0$-run of length~$z$}  
\KwOutput{$\enc[alg:rrfs-encode](\bfx)\in \bigcup_{m\leq n} \rrfs(m)$}
$j\leftarrow 1$ \\
\While{$j\leq \abs*{\bfx}-\ell'$}{
    \For{$i = j-1,\ldots,0$}{
    	\If{$\exists \bfy\in \ball[s]_t(\bfx)\colon \bfy_{i+[\ell']} = \bfy_{j+[\ell']}$} {
    		Replace $\bfx_{j+[\ell']}$ with $\bfs$ from~\eqref{eq:replaced} \\
    		$j\leftarrow \max\bracenv*{0,j-\ell'+1}$ \\
    		\Break
    	}
	}
	$j\leftarrow j+1$
}
\Return $\bfx$
\end{algorithm}
Our construction is summarized in \cref{alg:rrfs-encode}; its main 
idea is a generalization of \cite[Alg.~3]{EliGabMedYaa21}, as follows. 
Assume $a (1-H_q(\delta)) > 1$, and choose $\epsilon > 0$ such that 
$\zeta\deq a (1-H_q(\delta)-\epsilon) - 1 > 0$. Let $z = 
\floorenv*{\zeta \log_q(n)}$. An information string is first encoded 
into a length-$n$ string~$\bfx$ containing no $0$-run of length~$z$, 
which may be done in linear time using $\ceilenv[\big]{\frac{q}{q-2} 
n^{1-\zeta}} = O(n^{2-a\parenv*{1-H_q(\delta)-\epsilon}})$ redundant 
symbols \cite[Lem.~4]{YehBarMarYaa23}. Interestingly, this allows us 
to achieve redundancy which is arbitrarily close, in orders of 
magnitude, to the result of \cref{thm:ham}. Next, using 
\begin{align}
\ell'\deq 11 + \ceilenv*{2 \log_q(\ell) + a (1-\epsilon) \log_q(n)}
\leq \ell, 
\end{align}
where the last inequality holds for all sufficiently large~$n$, it is 
then iteratively checked whether $\bfx_{[\ell'+j]}\in 
\rrfs_{t, \ell'}(\ell'+j)$, for $j\in [n-\ell'+1]$ in increasing 
order.

If in some iteration it is determined that $\bfx_{[\ell'+j]}\not\in 
\rrfs_{t, \ell'}(\ell'+j)$, then the algorithm deletes~$\bfx_{j+
[\ell']}$ from~$\bfx$ and replaces it with a sequence with the 
following form:
\begin{align}\label{eq:replaced}
\bfs \deq 0^z 1 \circ E(j-i) \circ 1 0^{z'} 1 \circ E(\en(\bfe)) \circ 1,
\end{align}
where 
\begin{itemize}
\item 
$j,i$ are the loop-indices at any specific iteration of 
\cref{alg:rrfs-encode}, and by abuse of notation we take~$(j-i)$ to 
represent the $q$-ary expansion of the difference, using only as many 
symbols as required (since $j>i$, the all-zero representation $0^k$ 
would be taken to stand for $2^k$ instead of~$0$); 

\item 
$\bfe\deq \bfx_{j+[\ell']}-\bfx_{i+[\ell']}\in \Sigma^{\ell'}$ when 
$j-i\geq \ell'$, or $\bfe\deq \bfx_{i+[\ell'+k]}-\bfy_{i+[\ell'+k]}\in 
\Sigma^{\ell'+k}$ when $j-i<\ell'$; recall, however, that from the 
proof of \cref{lem:imp-pi} it follows that we may always assume 
$\abs*{\supp(\bfe)}\leq \floorenv*{3\ell'/2}$ is known, even for $k > 
\floorenv*{\ell'/2}$. In both cases $\wt(\bfe)\leq t$, and $\bfx_{j+
[\ell']}$ is recoverable from $\bfe, \bfx_{i+[j-i]}$; and

\item 
$z'\deq \ceilenv*{\log_q(\ell)}+2$ and~$E(\cdot)$ is the explicit and 
efficient encoder described in \cite[Alg.~1]{LevYaa19}; it can accept 
sequences of lengths at most~$q \ell$, and returns an encoded version 
containing no $0$-runs of length~$z'$, utilizing only a single 
redundant symbol. Observe that both $\log_q(j-i), \abs*{\en(\bfe)}\leq 
\frac{3}{2} \ell < q \ell$.
\end{itemize}

We note that since~$E(\cdot)$ may accept sequences of varying 
(sufficiently small) lengths, so too is~$\abs*{\bfs}$ not constant. 
Our next aim is to bound~$\abs*{\bfs}$ from above, as this affects the 
correct operation of \cref{alg:rrfs-encode}, namely, its termination 
condition. This behavior, and the operation of \cref{alg:rrfs-encode}, 
are demonstrated in the next example.

\begin{example}
A complete run of \cref{alg:rrfs-encode} will be tedious to track. 
We therefore limit ourselves to demonstrate a single step of the 
algorithm, for the follwing toy example: $q=2, n=16384, a=4, 
\delta=0.025$. Then, $\ell = 56, t = 1$. If we take $\epsilon = 0.41$, 
then we have $z = 9$. Also note $z' = 8$ and $\ell' = 56 = \ell$. 
Observe a sequence beginning with 
\begin{align*}
\bfx =\>& {\scriptstyle 01010111010101110101011101010111} \\
& {\scriptstyle 01010111010101110101011101011111 \ldots}
\end{align*}
which contains no run of zeros of length~$z=9$. 
Assume \cref{alg:rrfs-encode} reaches $j=8$ to consider this prefix 
of~$\bfx$; also observe that 
\begin{align*}
\bfy =\>& {\scriptstyle 01010111010101110101011101010111} \\
& {\scriptstyle 0101011101010111010101110101\underline{0}111 \ldots}
\end{align*}
satisfies $\bfy_{0+[\ell']} = \bfy_{0+[56]} = \bfy_{8+[56]} = \bfy_{j+
[\ell']}$. Therefore, \cref{alg:rrfs-encode} replaces $\bfy_{j+
[\ell']}$ with a substring $\bfs$ composed of 
\begin{align*}
\bfs \deq 0^9 1 \circ E(8) \circ 1 0^8 1 \circ E(\en(\bfe)) \circ 1,
\end{align*}
where 
\begin{itemize}
\item 
$8$ is represented by $000$, and $E(000) = 0001$ contains no zero-run 
of length~$z'=8$ and uses a single redundant symbol 
(we refrain from tracing \cite[Alg.~1]{LevYaa19} for brevity; suffice to 
note that $\abs*{E(000)} = 3+1 = 4$, that it contains no zero-run 
of length $z' = 8$, and that $000$ can be decoded from it).

\item 
$\bfe = \bfx_{0+[56+8]} - \bfy_{0+[56+8]} = 0^{60} 1 0^3\in 
\Sigma^{56+8}$, and its indexing in $\ball[s]_1(\bfzero)$ is $i(\bfe) = 
4$. That is, $\en(\bfe)\in \Sigma^{\ceilenv*{\frac{3}{2}\ell' H_q(2t/
3\ell')}} = \Sigma^8$ is represented $00000100$. Finally, $E(\en(\bfe)) 
= 000001001$ (again, we do not trace \cite[Alg.~1]{LevYaa19}).
\end{itemize}
Finally, 
\begin{align*}
\bfs = 0^9 1\circ 0001\circ 1 0^8 1 \circ 000001001 \circ 1,
\end{align*}
and its length is~$34 < 56 = \ell'$ (this fact is key to the algorithm's 
termination condition, as will be discussed next).
\end{example}

As a matter of convenience, we denote moving forward 
\begin{align}\label{eq:Hq}
\Hq(\delta) = 
H_q\parenv[\big]{\min\bracenv[\big]{\tfrac{q-1}{q}, \delta}}.
\end{align}

\begin{theorem}\label{thm:rrf-enc}
If $\frac{3}{2} \Hq(\frac{2}{3}\delta) - H_q(\delta)\leq \frac{1}{a}$, 
then \cref{alg:rrfs-encode} terminates, $\enc[alg:rrfs-encode](\bfx)
\in \bigcup_{m\leq n}\rrfs_{t,\ell}(m)$, and~$\bfx$ can be decoded 
from it.
\end{theorem}
\begin{IEEEproof}
First observe that if the last iteration of \cref{alg:rrfs-encode} 
terminates, then its output is resilient-repeat-free.

Next, we will show that the inserted substring in~\eqref{eq:replaced} 
is strictly less than~$\ell$, hence each replacement that the 
algorithm performs \emph{shortens}~$\bfx$. As a consequence, the 
algorithm must terminate. Indeed, observe that 
\begin{align*}
\abs*{\bfs} &= 8 + \floorenv*{\zeta \log_q(n)} + 
\ceilenv*{\log_q(j-i)} + \ceilenv*{\log_q(\ell)} + \abs*{\en(\bfe)} \\
&< 10 + \log_q(\ell) + \zeta \log_q(n) + \log_q(j-i) + 
\abs*{\en(\bfe)}.
\end{align*}
If $j-i\geq \ell'$, then we bound $\log(j-i)\leq \log(n)$, and we have 
seen that 
\begin{align*}
\abs*{\en(\bfe)} &\leq \ceilenv*{\ell' \Hq(t/\ell')} 
\leq \ceilenv*{\ell \Hq(t/\ell)} \\
&\leq \ceilenv*{\ell H_q(\delta)} 
< a H_q(\delta) \log_q(n) + 1, 
\end{align*}
hence in this case 
\begin{align*}
\abs*{\bfs} 
&< 11 + \log_q(\ell) + (\zeta + 1 + a H_q(\delta)) \log_q(n) \\
&= 11 + \log_q(\ell) + a (1-\epsilon) \log_q(n) 
< \ell', 
\end{align*}
as required. Otherwise we bound $\log_q(j-i)\leq \log_q(\ell')\leq 
\log_q(\ell)$ and 
\begin{align*}
\abs*{\en(\bfe)}
&\leq \ceilenv[\big]{(\ell'+k) \Hq(t/(\ell'+k))} \\
&< \tfrac{3}{2} \ell \Hq(\tfrac{2}{3} \delta) + 1, 
\end{align*}
where $k\deq \min\bracenv*{j-i, \floorenv*{\ell'/2}}$. Hence, 
\begin{IEEEeqnarray*}{+rCl+x*}
\abs*{\bfs} &<& 11 + 2 \log_q(\ell) + 
\parenv*{\zeta + \tfrac{3}{2} a \Hq(\tfrac{2}{3} \delta)} \log_q(n) \\
&\leq& \ell' + 
\parenv*{a (\tfrac{3}{2} \Hq(\tfrac{2}{3} \delta) - H_q(\delta)) - 1} 
\log_q(n).
\end{IEEEeqnarray*}
Under the assumption of the theorem, it also holds in this case that 
$\abs*{\bfs}<\ell'$.

Lastly, observe that, iterating over $j\in 
[\abs*{\enc[alg:rrfs-encode](\bfx)}-\ell']$ in decreasing order, the 
first observed instance of~$0^z1$ is always the last to have been 
inserted by \cref{alg:rrfs-encode}; this holds because after each 
replacement, $j$~is decreased only by~$\ell'-1$, hence any later 
replacements, say at index~$j'$, either satisfy $j'>j$ or they 
overwrite the first~$0$ of~$0^z1$ (observe that $\bfs$ ends with 
a~$1$). Further, by observing the first instance of $0^{z'}$ 
following that instance of~$0^z1$, it is possible to uniquely deduce 
the coordinates of~$E(j-i)$, and therefore to deduce~$i$. Now, given 
$E(\en(\bfe))$ one obtains~$\bfe$, and with $\bfe, \bfx_{i+[j-i]}$ is 
is uniquely possible to reconstruct the removed segment~$\bfx_{j+
[\ell]}$. Since every replacement of \cref{alg:rrfs-encode} is 
reversible, and the process can be tracked in reverse, $\bfx$ can be 
reconstructed.
\end{IEEEproof}

\begin{lemma}
The run-time of \cref{alg:rrfs-encode} is $O(n^2 \log(n)^2)$.
\end{lemma}
\begin{IEEEproof}
In any iteration, if there exists $\bfy\in 
\ball[s]_t(\bfx_{[\ell'+j]})\setminus\rf_{\ell'}(\ell'+j)$, and~$j$ is 
minimal such that this occurs, then there necessarily exists $i<j$ 
such that $\bfy_{i+[\ell']} = \bfy_{j+[\ell']}$. By 
\cref{lem:ham-dist}, if $i\leq j-\ell'$, the existence of such~$\bfy$ 
is equivalent to $\wt(\bfx_{j+[\ell']}-\bfx_{i+[\ell']})\leq t$, which 
may be verified in at most~$\ell' (j-\ell')$ operations. On the other 
hand, for each $0<k<\ell'$ we check whether there exists~$\bfy\in 
\ball[s]_t(\bfx_{[\ell'+j]})$ with~$i = j-k$; as seen in the proof of 
\cref{lem:imp-pi}, this implies that $\bfy_{i+[\ell'+k]}$ is 
$k$-periodic. The following procedure verifies whether such~$\bfy$ 
exists: denote for convenience $\bfu\deq \bfx_{i+[\ell'+k]}$; for each 
$p\in [k]$, we define the multiset $U_p\deq \multset*{u(q)}{q\equiv p 
\pmod k}$. If the most frequent element in~$U_p$ is some $x\in 
\Sigma$, denote by $t_p$ the number of occurrences of all other 
elements in $U_p$; clearly, there exists~$\bfy$ with the given~$i$ if 
and only if $\sum_{p\in [k]} t_p \leq t$. This algorithm requires at 
most~$O(\ell \log(\ell))$ operations for each~$k$ (due to the 
summation). For~$\ell = O(\log(n))$, any iteration requires at 
most~$O(n \log(n))$ operations in total, for both cases.

Finally, observe that there could be at most~$n \ell'\leq n \ell$ 
iterations of \cref{alg:rrfs-encode}, completing the proof.
\end{IEEEproof}

\subsection{Error-correcting codes for the noisy substring channel}

Based on \cref{cor:ham-rate}, we can demonstrate the existence of 
error-correcting codes for the noisy substring channel, which achieve 
at most a constant redundancy over that of classical error-correcting 
codes for Hamming noise.
\begin{corollary}\label{cor:encoding}
Let $C\subseteq\Sigma^n$ be an error-correcting code, capable of 
correcting $t$ substitution errors, and denote, for some $\bfz 
\in \Sigma^n$, $\bar{C}_{\bfz}\deq (\bfz+C) \cap \rrfs(n)$. 
Then for any $\bfx\in\bar{C}_{\bfz}$ and $\bfy\in \ball[s]_t(\bfx)$, 
it is possible to uniquely decode $\bfx$ observing only 
$Z_{\ell+1}(\bfy)$. 
Further, decoding is possible through a greedy algorithm for 
reconstruction of $\bfy$, followed by application of any decoding 
scheme for $C$.

Finally, in the cases indicated in~\cref{cor:ham-rate}, where 
$\red\parenv*{\rrfs(n)} = O(1)$, there exists~$\bfz$ satisfying 
$\red(\bar{C}_{\bfz}) = \red(C)+O(1)$.
\end{corollary}

Note that \cref{cor:encoding} is unfortunately nonconstructive. It is 
our hope that the encoder of \cref{alg:rrfs-encode} may be combined 
with error-correction techniques to yield explicit code constructions 
for this channel. However, achieving this goal seems to require new 
ideas, and we leave it for future study.

\section{Deletion noise}\label{sec:del}

This section is dedicated to the study of resilient-repeat-free 
sequences under deletion, rather than Hamming, errors. We demonstrate 
that the same probabilistic tools can be used to bound from above the 
redundancy of such sequences. We remark that the same method can be 
used to study insertion errors, even though the equivalence of 
insertion/deletion-correction does not extend in a straightforward 
manner to our setting.

For $\bfx\in \Sigma^n$, let $\sphere[d]_t(\bfx)\subseteq \Sigma^{n-t}$ 
denote the set of strings generated from~$\bfx$ by~$t$ deletions. 
Again, superscript $\mathrm{d}$ marks \emph{deletion} noise, and does 
not serve as a parameter.

\begin{definition}\label{def:rrf-del}
For integers~$t,\ell\leq n$, define a family of repeat-free strings 
which is resistant to 
deletion noise: 
\begin{align*}
\rrfd_{t,\ell}(n) &\deq \mathset*{\bfx\in\Sigma^n}{\sphere[d]_t(\bfx) 
\subseteq \rf_{\ell}(n-t)}.
\end{align*}
\end{definition}

Again, we fix $t,\ell$ as in \eqref{eq:tell}, and omit them from 
$\rrfd(n)$ whenever possible. Then we have the following:

\begin{theorem}\label{thm:del}
For all $a>1$ and $\delta>0$ it holds that 
\begin{align*}
\red\parenv*{\rrfd(n)} 
= O\parenv*{n^{2-a+\frac{2a(1+\delta)}{\log_2(q)} 
H_2\parenv*{\delta/(1+\delta)}} \Big/ \log(n)}.
\end{align*}
\end{theorem}
\begin{IEEEproof}
We follow a similar strategy as in \cref{thm:ham-red}, but apply a 
symmetric bound in \cref{cor:LLL}, i.e., utilizing a fixed constant 
$\phi_{I,J}\equiv \phi$. Note that a sufficient condition for $\bfx\in 
\rrfd(n)$ is that for every observable pair $(I,J)\in 
\binom{[n]}{\ell_a}^2$, such that 
\begin{align}\label{eq:del-obsrv}
I(\ell-1)-I(0) < \ell + t
\end{align}
(and similarly for $J$), it holds that $\bfx_I\neq \bfx_J$. 
For such a pair, denote 
\begin{align*}
A_{I,J} \deq \mathset*{\bfx\in \Sigma^n}{\bfx_I = \bfx_J} 
= \mathset*{\bfx\in\Sigma^n}{\bfu_{I,J}=0}.
\end{align*}
Again, we let $\bfx\in\Sigma^n$ be chosen uniformly at random, 
implying $\red\parenv[\big]{\rrfd(n)} = -\log_q 
\Pr\parenv[\big]{\bfx\in\rrfd(n)}$.

In order to apply \cref{cor:LLL} we need to determine the 
constant~$\phi$, the neighborhoods $\Gamma_{I,J}$ (establishing an 
independence condition) and their sizes. 
For any observable pair $(I,J)$, note that $\Pr\parenv*{\bfx\in 
A_{I,J}} = q^{-\ell} \leq q\cdot n^{-a}$, and for convenience denote 
$\pid\deq q\cdot n^{-a}$ and $\phi\deq e \pid$. 
Next, by \cref{lem:ind} it suffices that $\Gamma_{I,J}$ consists of 
all $(P,Q)\in \Gamma_I$ satisfying \eqref{eq:del-obsrv}. Thus, to 
determine~$P$, it suffices to choose 
\begin{enumerate}
\item
a single element of $I$ (which shall be a member of $P\cap I$);

\item
an interval of length $\ell+t$ containing the chosen element; and

\item
any $\ell-1<\ell$ additional elements of the chosen interval.
\end{enumerate}
Then $Q$ can be chosen from any interval of length $\ell+t$. 
The same holds for a suitable choice of $Q\cap I\neq\emptyset$. Thus, 
$
\abs*{\Gamma_{I,J}} \leq \ell (\ell+t) n \binom{\ell+t}{\ell}^2
$.

Now, in order to satisfy the conditions of \cref{cor:LLL} we observe 
from $\binom{s}{r}\leq \sqrt{\frac{s}{r (s-r)}} 2^{s H_2(r/s)}$ (a 
relaxation of, e.g., \cite[Ch.10,~Sec.11,~Lem.7]{MacSlo78}) that 
\begin{align*}
\binom{\ell+t}{t}^2\leq 
\frac{\ell+t}{\ell t}
n^{\frac{2a(1+\delta)}{\log_2(q)} H_2\parenv*{\delta/(1+\delta)}}.
\end{align*}
If $\frac{2(1+\delta)}{\log_2(q)} H_2\parenv[\big]{\frac{\delta}{1+
\delta}} \geq 1$ or $a < \parenv[\big]{1-\frac{2(1+\delta)}{\log_2(q)} 
H_2\parenv[\big]{\frac{\delta}{1+\delta}}}^{-1}$, the theorem 
vacuously holds. Otherwise, we note that 
\begin{align*}
(\abs*{\Gamma_{I,J}}+1) \pid &\leq q n^{-a} + q 
\frac{(\ell+t)^2}{t} n^{1-a + \frac{2a(1+\delta)}{\log_2(q)} 
H_2\parenv*{\frac{\delta}{1+\delta}}} \\
&= o_n(1).
\end{align*}
Then, we observe 
\begin{align*}
\phi_{I,J} \exp\parenv[\Big]{-\sum_{\mathclap{(P,Q)\in 
\Gamma_{I,J}}} \phi_{P,Q} - \phi_{I,J}} &= 
\pid e^{1 - e (\abs*{\Gamma_{I,J}}+1) \pid} \\
&> \pid \geq \Pr\parenv*{\bfx\in A_{I,J}}.
\end{align*}

Finally, one needs also note that the number of observable pairs 
$(I,J)$ satisfying \cref{eq:del-obsrv} is no more than 
$\binom{n-\ell-t}{2}\cdot \binom{\ell+t}{\ell}^2 < 
n^2 \binom{\ell+t}{\ell}^2$. From \cref{cor:LLL} it follows that 
\begin{align*}
\Pr\parenv[\bigg]{\bfx\not\in\bigcup_{I,J} A_{I,J}} 
&\geq 
\exp\parenv*{-e\pid n^2 \binom{\ell+t}{\ell}^2}, 
\end{align*}
and hence 
\begin{align*}
\red\parenv*{\rrfd(n)} &= -\log_q \Pr\parenv*{x\in\rrfd(n)} \\
&\leq e\log(e) \pid n^2 \binom{\ell+t}{\ell}^2 \\
&\leq e\log(e) q \frac{\ell+t}{\ell t} n^{2 - a + 
\frac{2a(1+\delta)}{\log_2(q)} H_2\parenv*{\delta/(1+\delta)}}, 
\end{align*}
which completes the proof.
\end{IEEEproof}

\begin{corollary}
If 
$
a > \parenv[\big]{1 -\frac{2(1+\delta)}{\log_2(q)} H_2\parenv[\big]{\frac{\delta}{1+\delta}}}^{-1}
$, 
for any $\delta>0$, then $R(\rrfd(n)) = 1-o_n(1)$, and if 
$
a\geq 2 \parenv[\big]{1 -\frac{2(1+\delta)}{\log_2(q)} 
H_2\parenv[\big]{\frac{\delta}{1+\delta}}}^{-1}
$ 
then $\red\parenv*{\rrfd(n)}=O_n(1)$. 
\end{corollary}

Note again that if $a>1$ (respectively $a>2$), then for all 
sufficiently small $\delta>0$ it holds that $R\parenv{\rrfd(n)} = 
1-o_n(1)$ (respectively, $\red\parenv[\big]{\rrfd(n)} = O_n(1)$). 
Before concluding, we also note that a parallel statement to 
\cref{cor:encoding} holds in this setting, as well.

\section{Secondary structure avoidance}\label{sec:secstruct}

\begin{figure}[t]
\begin{center}
\begin{tikzpicture}[scale=0.8,transform shape,auto,node distance=3,
    main node/.style={thick,circle,draw,font=\sffamily\large}]

  \node[style={font=\sffamily\large}] (fd3) at (-1.837,-1.634) {$3'$};
  \node[style={font=\sffamily\large}] (fu3) at (-2.24,1.95) {$5'$};

  \node[main node] (fd2) at (-1.2,-0.88) {$A$};
  \node[main node] (fu2) at (-1.42,1.6) {$A$};
  \path (fd3) edge[-] (fd2);
  \path (fu3) edge[-] (fu2);

  \node[main node] (fd1) at (-0.71,-0.3) {$G$};
  \node[main node] (fu1) at (-0.71,1.3) {$T$};

  \node[main node] (sd1) at (0,0) {$T$};
  \node[main node] (su1) at (0,1) {$A$};
  \path (sd1.north) edge[double] (su1.south);

  \node[main node] (sd2) at (0.77,0) {$C$};
  \node[main node] (su2) at (0.77,1) {$G$};
  \path (sd2.north) edge[double distance=2pt] (su2.south);
  \path (sd2.north) edge[] (su2.south);
  
  \node[main node] (sd3) at (1.54,0) {$T$};
  \node[main node] (su3) at (1.54,1) {$A$};
  \path (sd3.north) edge[double] (su3.south);
  
  \node[main node] (sd4) at (2.31,0) {$G$};
  \node[main node] (su4) at (2.31,1) {$C$};
  \path (sd4.north) edge[double distance=2pt] (su4.south);
  \path (sd4.north) edge[] (su4.south);
  
  \node[main node] (sd5) at (3.08,0) {$A$};
  \node[main node] (su5) at (3.08,1) {$T$};
  \path (sd5.north) edge[double] (su5.south);
  
  \node[main node] (sd6) at (3.85,0) {$G$};
  \node[main node] (su6) at (3.85,1) {$C$};
  \path (sd6.north) edge[double distance=2pt] (su6.south);
  \path (sd6.north) edge[] (su6.south);
  
  \node[main node] (sd7) at (4.62,0) {$C$};
  \node[main node] (su7) at (4.62,1) {$G$};
  \path (sd7.north) edge[double distance=2pt] (su7.south);
  \path (sd7.north) edge[] (su7.south);
  
  \node[main node] (l1) at ([shift=(262:0.92)] 5.39,0.5) {$G$};
  
  \node[main node] (l2) at ([shift=(311:0.92)] 5.39,0.5) {$C$};
  
  \node[main node] (l3) at ([shift=(0:0.92)] 5.39,0.5) {$T$};
  
  \node[main node] (l4) at ([shift=(49:0.92)] 5.39,0.5) {$A$};
  
  \node[main node] (l5) at ([shift=(98:0.92)] 5.39,0.5) {$A$};
  
  
  \path (sd1.south west)
    edge[decorate,decoration={brace,mirror,raise=4.5},"stem (length$\>=7$)"below=0.2]
        (sd1.south west -| l1.west);
  
  \path (l1.south west)
    edge[decorate,decoration={brace,mirror,raise=4.5},"loop"below=0.2]
        (l1.south west -| l3.east);

\end{tikzpicture}
\end{center}
\caption{Formation of a hairpin-loop secondary structure in an oligonucleotide.\label{fig:haripin}}
\end{figure}
\begin{figure}[t]
\begin{center}
\begin{tikzpicture}[scale=0.8,transform shape,auto,node distance=3,
    main node/.style={thick,circle,draw,font=\sffamily\large}]

  \node[style={font=\sffamily\large}] (fd3) at (-1.837,-1.634) {$3'$};
  \node[style={font=\sffamily\large}] (fu3) at (-2.24,1.95) {$5'$};

  \node[main node] (fd2) at (-1.2,-0.88) {$A$};
  \node[main node] (fu2) at (-1.42,1.6) {$A$};
  \path (fd3) edge[-] (fd2);
  \path (fu3) edge[-] (fu2);

  \node[main node] (fd1) at (-0.71,-0.3) {$G$};
  \node[main node] (fu1) at (-0.71,1.3) {$T$};

  \node[main node] (sd1) at (0,0) {$T$};
  \node[main node] (su1) at (0,1) {$A$};
  \path (sd1.north) edge[double] (su1.south);

  \node[main node] (sd2) at (0.73,-0.2) {$G$};
  \node[main node] (su2) at (0.73,1.2) {$G$};
  
  \node[main node] (sd3) at (1.46,0) {$T$};
  \node[main node] (su3) at (1.46,1) {$A$};
  \path (sd3.north) edge[double] (su3.south);
  
  \node[main node] (sd4) at (2.23,0) {$G$};
  \node[main node] (su4) at (2.23,1) {$C$};
  \path (sd4.north) edge[double distance=2pt] (su4.south);
  \path (sd4.north) edge[] (su4.south);
  
  \node[main node] (sd5) at (2.97,-0.2) {$G$};
  \node[main node] (su5) at (2.97,1.2) {$T$};
  
  \node[main node] (sd6) at (3.73,-0.2) {$A$};
  \node[main node] (su6) at (3.73,1.2) {$C$};
  
  \node[main node] (sd7) at (4.47,0) {$C$};
  \node[main node] (su7) at (4.47,1) {$G$};
  \path (sd7.north) edge[double distance=2pt] (su7.south);
  \path (sd7.north) edge[] (su7.south);
  
  \node[main node] (l1) at ([shift=(262:0.92)] 5.24,0.5) {$G$};
  
  \node[main node] (l2) at ([shift=(311:0.92)] 5.24,0.5) {$C$};
  
  \node[main node] (l3) at ([shift=(0:0.92)] 5.24,0.5) {$T$};
  
  \node[main node] (l4) at ([shift=(49:0.92)] 5.24,0.5) {$A$};
  
  \node[main node] (l5) at ([shift=(98:0.92)] 5.24,0.5) {$A$};
  
  
  \path (sd1.south west)
    edge[decorate,decoration={brace,mirror,raise=9},"stem (distance$\>=3/7$)"below=0.4]
        (sd1.south west -| l1.west);
  
\end{tikzpicture}
\end{center}
\caption{Imperfect stem in a hairpin-loop structure.\label{fig:haripin-dist}}
\end{figure}

In this section, we leverage \cref{alg:rrfs-encode} to protect against 
the formation of secondary structures in coded DNA strands. We focus 
on a special type of secondary structure, called \emph{hairpin-loop} 
(see \cref{fig:haripin}). Unlike recent works, our analysis does not 
require a perfect binding in the stem region of the hairpin structures 
(see \cref{fig:haripin-dist}), and we show that \cref{alg:rrfs-encode} 
can be utilized to avoid the formation of such structures. However, we 
rely in this section on the Hamming metric rather than the Levenshtein 
metric as was suggested in~\cite{MilKas06}, thus we do not consider 
the formation of so-called \emph{bulge-loops} due to the elasticity of 
the DNA sugar-phosphate backbone. 
We remark that given an efficient enumeration of the Levenshtein ball 
about any point, these methods can be extended to utilize that metric, 
too.

In order to define hairpin-loop-avoiding sequences, we first present 
some notation. An involution on~$\Sigma$ is a mapping $x\mapsto 
\bar{x}$ such that for all~$x\in \Sigma$ it holds that 
$\overline{\bar{x}} = x$; we now assume~$\Sigma$ to be equipped with 
such an involution (we allow fixed points, in order to account for 
odd~$q$, which shall not affect our analysis). For example, DNA is 
composed of four \emph{nucleotide bases}: the \emph{purines}, 
adenine~($\bfA$) and guanine~($\bfG$), are respectively the 
complements of the \emph{pyrimidines} thymine~($\bfT$) and 
cytosine~($\bfC$); when forming a \emph{double helix} (or 
\emph{duplex}) structure, each base can only stably bond 
(\emph{hybridize}) with its complement. For $\bfx \in \Sigma^n$, 
denote $\bar{\bfx}\deq \bar{x}(0) \bar{x}(1) \cdots \bar{x}(n-1)$.

DNA strands are also oriented: each nucleotide is composed of one of 
four nitrogenous bases, together with a pentose sugar and a phosphate 
group; the phosphate groups connect the sugar rings of adjacent 
nucleotides 5'-end to 3'-end (referring to the five-carbon sites of 
the sugar rings) to form a long chain (\emph{oligonucleotide}), and 
thus the orientation can be observed from any segment of the chain. 
Stable duplexes only form between oligonucleotides of reverse 
orientations, and therefore coiled-loop secondary structures cannot 
appear. To capture this notion, we denote for~$\bfx\in \Sigma^n$ the 
\emph{reverse} sequence $\bfx\rv\deq x(n-1) \cdots x(1) x(0)$.

For integers $t\leq \ell$, we define the set of length-$n$
\emph{$(t,\ell)$-hairpin avoiding} strings to contain those strings that do not have the potential for the formation of a loop with 
stem-length~$\ell$, of which at least $\ell-t$ symbols are hybridized. More precisely, 
\begin{IEEEeqnarray*}{+rCl+x*}
\ha_{t,\ell}(n) &\deq& \mathset*{\bfx\in \Sigma^n}{\scriptsize\begin{matrix}
\forall 0\leq i<j < n \\
\forall \ell-t\leq \ell'\leq \min\bracenv*{\ell, j-i, n-j} \\
\ds(\bfx_{i+[\ell']}, (\bar{\bfx}_{j+[\ell']})\rv) 
> t - (\ell-\ell')
\end{matrix}}.
\end{IEEEeqnarray*}
Observe in the above definition, that for $0<i<j<n-1$, 
\begin{align*}
& \ds(\bfx_{i-1+[\ell'+1]}, (\bar{\bfx}_{j+[\ell'+1]})\rv) > t - (\ell-\ell') + 1 \\
& \implies 
\ds(\bfx_{i+[\ell']}, (\bar{\bfx}_{j+[\ell']})\rv) > t - (\ell-\ell'), 
\end{align*}
hence some conditions in the above definition are redundant.

As before, for fixed real numbers $a>1$ and $0<\delta<1$, we also make 
the notation $\ha_{\delta,a}(n)$. 
We will show that when $a > (1-\Hq(\delta))^{-1}$ 
(for $\delta < 1-\frac{1}{q}$) then \cref{alg:rrfs-encode} can, with 
slight necessary adjustments, encode into $\bigcup_{m\leq n}
\ha_{t_\delta,\ell_a}(m)$ with redundancy $O(n^{2-a\parenv*{1-
H_q(\delta)-\epsilon}})$ for arbitrarily small $\epsilon>0$. 
We leave the interesting problem of stating an analogue of 
\cref{lem:ham-conv} in this case for future study.

\begin{algorithm}[t]
\caption{Hairpin-avoiding Encoder}
\label{alg:hairpin-encode}
\SetKwInput{KwInput}{Input}                
\SetKwInput{KwOutput}{Output}  
\SetAlgoLined
\KwInput{ $\bfx\in \Sigma^n$ containing no $0$-run of length~$z$}  
\KwOutput{$\enc[alg:rrfs-encode](\bfx)\in \bigcup_{m\leq n} \rrfs_{t_\delta,\ell_a}(m)$}
$j'\leftarrow 2 (\ell_a-t_\delta)$ \\
\While{$j'\leq \abs*{\bfx}$}{
	\For{$i = j' - 2 (\ell_a-t_\delta),\ldots,0$}{
		$\ell'\leftarrow \min\bracenv[\big]{\ell_a, \floorenv[\big]{\frac{j'-i}{2}}}$ \\
		$j\leftarrow j'- \ell'$ \\
		\If{$d\parenv*{(\bar{\bfx}_{j+[\ell']})\rv, \bfx_{i+[\ell']}} \leq t_\delta - (\ell_a-\ell')$} {
    		Replace $\bfx_{j+[\ell']}$ with $\bfs$ from~\eqref{eq:replaced-hairpin} \\
    		$j'\leftarrow \max\bracenv*{2 (\ell_a-t_\delta), j'-\ell'+1}$ \\
    		\Break
    	}
	}
	$j'\leftarrow j'+1$
}
\Return $\bfx$
\end{algorithm}

Indeed, the encoder presented in \cref{alg:hairpin-encode} differs 
from \cref{alg:rrfs-encode} only in the type of condition in the inner 
loop (ranges for $j,i$ are adjusted accordingly); if a replacement is 
required, instead of necessarily replacing an entire $\ell$-substring, 
in the case $i>j-\ell$ (i.e., $\ell'<\ell$) only the $\ell'$-suffix is 
replaced, with the substring 
\begin{align}\label{eq:replaced-hairpin}
\bfs \deq 0^z 1 \circ E(j-i) \circ 1 0^{z'} 1 \circ E(\en(\bfe)) \circ 1.
\end{align}
For convenience we repeat the previous definitions for the following 
expressions:
\begin{itemize}
\item 
$\zeta = a (1-H_q(\delta)-\epsilon) - 1 > 0$;

\item 
$z = \floorenv*{\zeta \log_q(n)}$;

\item 
$j-i$ represents the $q$-ary expansion of the difference (using only 
as many symbols as required), 
\end{itemize}
and we adjust the following definitions: 
\begin{itemize}
\item 
$\bfe\deq (\bar{\bfx}_{j+[\ell']})\rv-\bfx_{i+[\ell']}$;

\item 
$z'\deq \ceilenv*{\log_q(\ell)}+1$; and, finally, 

\item 
$E(\cdot)$ is an explicit and efficient encoder into strings 
containing no $0$-runs of length~$z'$, accepting inputs of lengths at 
most~$\ell$ and requiring a single redundant symbol 
\cite[Alg.~1]{LevYaa19}. We shall see below that both $\log(j-i), 
\abs*{\en(\bfe)}\leq \ell$.
\end{itemize}

The analysis of \cref{alg:hairpin-encode} is much similar to that of 
\cref{alg:rrfs-encode} in \cref{sec:rrf-enc}. We summarize the result 
in the following theorem.

\begin{theorem}\label{thm:ha-enc}
If $(1-\Hq(\delta))^{-1} < a < \delta^{-1}(1-\Hq(\delta))^{-1}$, 
then for sufficiently large~$n$ \cref{alg:hairpin-encode} terminates, 
$\enc[alg:hairpin-encode](\bfx)\in \bigcup_{m\leq n}\ha_{t,\ell}(m)$, 
and~$\bfx$ can be decoded from it.
\end{theorem}
\begin{IEEEproof}
We prove only the first part; the latter two follow exactly as in the 
proof of \cref{thm:rrf-enc}. As before, 
\begin{align*}
\abs*{\bfs} 
&< 
9 + \log(\ell) + \zeta \log_q(n) + \log_q(j-i) + \abs*{\en(\bfe)}.
\end{align*}
Repeating the analysis of \cref{thm:rrf-enc}, if $j-i\geq \ell$ (i.e., 
$\ell' = \ell$), then (bounding $\log_q(j-i)\leq \log_q(n)$) we have 
\begin{align*}
\abs*{\en(\bfe)} &\leq \ceilenv*{\ell \Hq(t/\ell)} 
\leq \ceilenv*{\ell_a H_q(\delta)} \\
&< a H_q(\delta) \log_q(n) + 1, 
\end{align*}
as before, and hence again (for sufficiently large~$n$) 
\begin{align*}
\abs*{\bfs} 
&< 9 + \log_q(\ell) + (\zeta + 1 + a H_q(\delta)) \log_q(n) \\
&= 9 + \log_q(\ell) + a (1-\epsilon) \log_q(n) < \ell.
\end{align*}
It follows that such an iteration of \cref{alg:hairpin-encode} also 
shortens~$\bfx$.

Otherwise, when $\ell'<\ell$ we again bound $\log_q(j-i)\leq 
\log_q(\ell') < \log_q(\ell)$ and 
\begin{align*}
\abs*{\en(\bfe)}
&\leq \ceilenv[\big]{\ell' \Hq\parenv[\big]{\tfrac{t-(\ell-\ell')}{\ell'}}} \\
&\leq \ell' \Hq\parenv[\big]{\tfrac{\ell' - (1-\delta) \ell}{\ell'}} + 1 \\
&< \ell' H_q(\delta) + 1. 
\end{align*}
Hence, 
\begin{IEEEeqnarray*}{+rCl+x*}
\abs*{\bfs} &<& 10 + 2 \log_q(\ell) + \zeta \log_q(n) + 
H_q(\delta) \ell' \\
&=& 10 + 2 \log_q(\ell) - (\epsilon a + 1) \log_q(n) \\*
&& +\> (1-H_q(\delta)) a \log_q(n) + H_q(\delta) \ell' \\
&\leq& 11 + 2 \log_q(\ell) - (\epsilon a + 1) \log_q(n) \\*
&& +\> (1-H_q(\delta)) \ell + H_q(\delta) \ell' \\
&\leq& 11 + 2 \log_q(\ell) - (\epsilon a + 1) \log_q(n) \\*
&& +\> (1-H_q(\delta)) (\ell-\ell') + \ell' \\
&\leq& (11 + 2 \log_q(\ell) - \epsilon a \log_q(n)) + \ell' \\*
&& +\> \parenv*{(1-H_q(\delta)) \delta a - 1} \log_q(n) 
< \ell', 
\end{IEEEeqnarray*}
where the last inequality again holds for sufficiently large~$n$, and 
relies on the theorem's assumption.
\end{IEEEproof}

\begin{corollary}
For all $a>1$ and sufficiently small~$\delta>0$, there exists an 
efficient (explicit) encoder from $\Sigma^{n-o(n)}$ into 
$\bigcup_{m\leq n} \ha_{t,\ell}(m)$, for sufficiently large~$n$.
\end{corollary}

The problems of encoding directly into $\ha_{t,\ell}(n)$, more 
precisely bounding its redundancy, as well as generalization to the 
Levenshtein metric (hence, considering also the formation of 
bulge-loop secondary structures), are left for future study.

\section*{Acknowledgments}

The authors gratefully acknowledge the invaluable insights provided by 
the anonymous reviewers and associate editor, which were instrumental 
in improving the presentation of the paper.

\bibliographystyle{IEEEtran}


\begin{IEEEbiography}[{\includegraphics[width=1in,height=1.25in,clip,keepaspectratio]{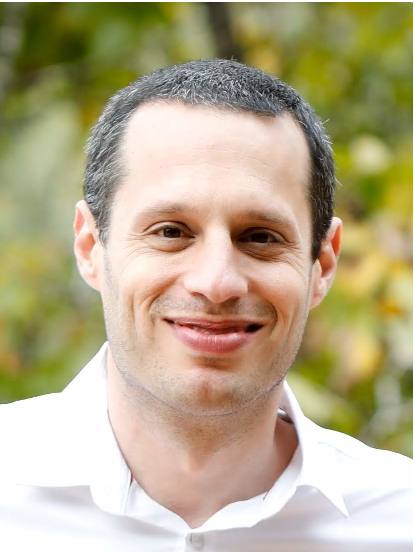}}]
{Yonatan Yehezkeally}
(S'12--M'20)
received the B.Sc.~degree (\emph{cum laude}) in mathematics and the M.Sc.~(\emph{summa cum laude}) and Ph.D. degrees in electrical and computer engineering from Ben-Gurion University of the Negev, Beer-Sheva, Israel, in~2013, 2017, and~2020 respectively. 
He is currently a Carl Friedrich von Siemens Post-Doctoral Research Fellow of the Alexander von Humboldt Foundation, with the Associate Professorship of Coding and Cryptography (Prof. Wachter-Zeh), School of Computation, Information and Technology, Technical University of Munich. 
His research interests include coding theory and algorithms, particularly with applications to novel storage media, with a focus on DNA-based storage and nascent sequencing technologies. They further include combinatorial analysis and structures, as well as algebraic structures.
\end{IEEEbiography}

\begin{IEEEbiography}[{\includegraphics[width=1in,height=1.25in,clip,keepaspectratio]{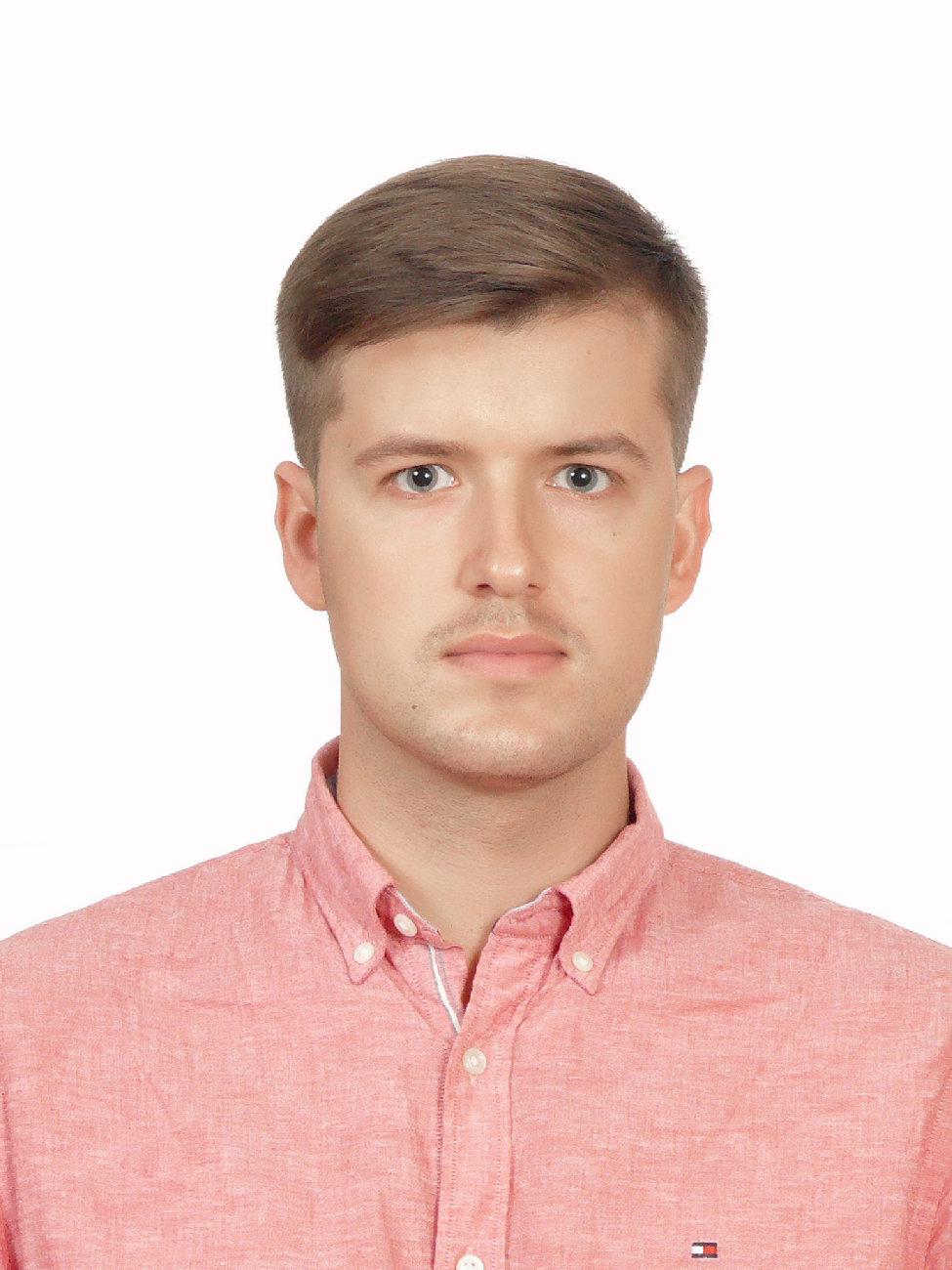}}]
{Nikita Polyanskii}
(Member, IEEE)
received the Specialist~degree in mathematics and the Ph.D.~degree in mathematics from the Lomonosov Moscow State University, Russia, in~2013 and~2016, respectively. 
From~2015 to~2017, he was a Researcher with the Institute for Information Transmission Problems, Russia, and a Senior Research Engineer with the Huawei Moscow Research Center, Russia. From~2017 to~2021, he was a Post-Doctoral Researcher with the Technion---Israel Institute of Technology, the Skolkovo Institute of Science and Technology, and the Technical University of Munich. He is currently a Senior Research Engineer with the IOTA Foundation, Germany. 
His research interests include the theory of error-correcting codes and distributed ledger technologies.
\end{IEEEbiography}

\end{document}